\documentclass[aps,prl,a4paper,10pt,twocolumn,superscriptaddress,floatfix,longbibliography,nofootinbib,nobalancelastpage]{revtex4-2}

\usepackage{docmute}
\newcommand{\combinedarxiv}{}
\usepackage[english]{babel}
\usepackage[utf8]{inputenc}
\usepackage[T1]{fontenc}
\usepackage[nice]{nicefrac}
\usepackage{graphicx}
\usepackage[dvipsnames]{xcolor}
\usepackage{amsmath,amssymb,bm,bbm}
\usepackage{braket}
\usepackage[normalem]{ulem}                         % strikethrough
\usepackage{orcidlink}
\usepackage{physics}
\usepackage{booktabs}
\usepackage{hyperref}
\usepackage{xspace}
\usepackage{parskip}
\newcommand{\maineqref}[1]{\eqref{#1}}
\newcommand{\mainref}[1]{\ref{#1}}

\newcommand{\btheta}{\boldsymbol\theta}

\newcommand{\inner}[2]{\langle #1 | #2 \rangle}
\newcommand{\expect}[3]{\langle #1 | #2 | #3 \rangle}
\newcommand{\avg}[1]{\langle #1 \rangle}            % expectation <A>
\newcommand{\Rez}{\mathrm{Re}}

\newcommand{\ie}{{\it i.e.},\ }
\newcommand{\eg}{{\it e.g.},\ }
\newcommand{\etal}{{\it et al.}\ }

\newcommand{\impl}{\textit{implicit}\xspace}
\newcommand{\expl}{\textit{explicit}\xspace}

\newcommand{\Expl}{\textit{Explicit}\xspace}

\definecolor{fedor}{rgb}{0.5, 0.0, 0.5}

\hypersetup{colorlinks, linkcolor=BrickRed, citecolor=MidnightBlue, urlcolor=blue}

\makeatletter
\g@addto@macro\frontmatter@above@affiliation@script{\parskip\z@}
\makeatother

\begin{document}

% ---------------------------------------------------------------------------
% Metadata
% ---------------------------------------------------------------------------
\title{Augmented Imaginary-Time Evolution of Neural Quantum States \\ with Implicit Time Steps}

\author{Maksymilian Kliczkowski\orcidlink{0000-0002-7987-9913}}
\affiliation{Center for Advanced Systems Understanding (CASUS), 02826 G\"orlitz, Germany\looseness=-3}
\affiliation{Helmholtz-Zentrum Dresden-Rossendorf (HZDR), 01328 Dresden, Germany\looseness=-3}

\author{Thomas D. K\"uhne}
\affiliation{Center for Advanced Systems Understanding (CASUS), 02826 G\"orlitz, Germany\looseness=-3}
\affiliation{Helmholtz-Zentrum Dresden-Rossendorf (HZDR), 01328 Dresden, Germany\looseness=-3}
\affiliation{Institute of Artificial Intelligence, Technische Universit\"at Dresden, Helmholtzstra{\ss}e 10, 01069 Dresden, Germany\looseness=-3}

\author{Fedor  Šimkovic IV}
\affiliation{IQM Quantum Computers, Georg-Brauchle-Ring 23-25, 80992 Munich, Germany}

\author{Prachi Sharma}
\affiliation{IQM Quantum Computers, Georg-Brauchle-Ring 23-25, 80992 Munich, Germany}

\author{Carlos L. Benavides-Riveros\orcidlink{0000-0001-6924-727X}}
\affiliation{IQM Quantum Computers, Georg-Brauchle-Ring 23-25, 80992 Munich, Germany}

\author{Werner Dobrautz\orcidlink{0000-0001-6479-1874}}
\affiliation{Center for Advanced Systems Understanding (CASUS), 02826 G\"orlitz, Germany\looseness=-3}
\affiliation{Helmholtz-Zentrum Dresden-Rossendorf (HZDR), 01328 Dresden, Germany\looseness=-3}
\affiliation{Center for Scalable Data Analytics and Artificial Intelligence (ScaDS.AI) Dresden/Leipzig, 01069 Dresden, Germany}
\affiliation{Technical University Dresden, 01069 Dresden, Germany}

\date{\today}

\begin{abstract}
Finding the ground states of quantum many-body systems remains a central challenge in physics and quantum chemistry. Imaginary-time evolution (ITE) offers a natural, widely used route, but its convergence is typically asymptotic. Variational approaches restrict this evolution to parameterized manifolds, with neural quantum states (NQS) providing flexible representations commonly optimized via stochastic reconfiguration (SR). Their performance depends on the geometry of the variational manifold, the convergence properties of ITE, and the stability of finite-step \expl (forward) integrators. By exploiting local geometric information of the energy landscape, the recently introduced augmented imaginary-time evolution (AITE) can, in principle, reach the ground state in finite imaginary time. To exploit the accelerated convergence of AITE in NQS optimization, we first combine an \impl Euler step with Galerkin projection to derive an \impl SR scheme that permits large, stable steps and substantially outperforms conventional methods. We use it to formulate an \impl version of AITE and show that its acceleration persists under variational projection and finite-step integration with asymptotically quadratic convergence of the energy error. With finite sampling, \impl schemes yield consistent gains, and \impl AITE becomes increasingly advantageous.
\end{abstract}

\keywords{neural quantum states; imaginary-time evolution; variational Monte Carlo; stochastic reconfiguration; imaginary-dynamics;backward Euler; implicit; AITE; ITE}

\maketitle

% ============================================================================
% \section{Introduction}\label{sec:intro}
% ===========================================================================

% ---------------------------------------
% 1) General Intro
% ---------------------------------------
The ground state of a quantum many-body system captures its zero-temperature properties, yet the exponential growth of the Hilbert space limits the exact diagonalization to small system sizes~\cite{lin_exactdiagonalizationmethods_1993}. To avoid this \emph{curse of dimensionality}, variational approaches restrict the wavefunction to a parameterized manifold far smaller than the associated Hilbert space~\cite{hackl_geometryvariationalmethods_2020,wu_variationalbenchmarks_2024}. Tensor networks represent variational wavefunctions~\cite{white_densitymatrixformulation_1992,schollwock_densitymatrixrenormalizationgroup_2011,orus_tensornetworkscomplex_2019}, and variational Monte Carlo (VMC) optimizes them by stochastic sampling~\cite{foulkes_quantummontecarlo_2001,becca_quantummontecarlo_2017}. Both exploit that low-energy states of many physical systems are well approximated by compact representations~\cite{verstraete_matrixproductstates_2006, hastings_arealawone_2007}. Neural quantum states (NQS) use neural networks (NN) for parametrization~\cite{carleo_solvingquantummanybody_2017,torlai_neuralnetworkquantumstate_2018,schmitt_quantummanybodydynamics_2020,hibat_allah_recurrentneuralnetwork_2020,reh_optimizingdesignchoices_2023}. Combined with VMC, NQS have been successfully applied across various interacting quantum many-body systems, as reviewed in Refs.~\cite{medvidovic_neuralnetworkquantumstates_2024,lange_architecturesapplicationsreview_2024}. Nevertheless, accurate ground-state optimization remains a challenge for large and complex systems, particularly in more than one dimension~\cite{carleo_solvingquantummanybody_2017,choo_studytwodimensionalfrustrated_2019,pfau_abinitiosolutionmanyelectron_2020,reh_optimizingdesignchoices_2023}.

% ----------------------------------------
% 2) ITE
% ----------------------------------------

Given a variational manifold, the goal is to efficiently evolve the state toward the ground state. A natural route is provided by imaginary-time evolution (ITE), formally related to real-time evolution through the Wick rotation $t\rightarrow-i\tau$, which replaces unitary dynamics by a non-unitary propagator~\cite{wick_propertiesbethesalpeterwave_1954,mclachlan_variationalsolutiontimedependent_1964,mcardle_variationalansatzbasedquantum_2019a,PhysRevLett.93.207204}. ITE underlies projector quantum Monte Carlo~\cite{zhang_constrainedpathmonte_1997,sandvik_loopupdatesvariational_2010,booth_fermionmontecarlo_2009,cleland_communicationssurvivalfittest_2010,spencer_handeqmcproject_2019,guther_necinelectronconfiguration_2020,shi_recentdevelopmentsauxiliaryfield_2021}
and is widely used to optimize NQS towards ground states~\cite{carleo_solvingquantummanybody_2017,ledinauskas_scalableimaginarytime_2023}. 
%It also provides a starting point for quantum algorithms for preparing ground and thermal states \cite{PhysRevA.111.012424}. 
For Hamiltonian $\hat{H}$ %with eigenenergies $E_0<E_1<\cdots<E_{\max}$, 
and an initial state $\ket{\psi(0)}$ with nonzero ground-state overlap, evolution $\smash{\ket{\psi(\tau)}=e^{-\tau\hat H}\ket{\psi(0)}}$ exponentially suppresses excited-state contributions, reaching the ground state asymptotically as $\tau\rightarrow\infty$. For the normalized state $\smash{\ket{\phi(\tau)}=\ket{\psi(\tau)}/\lVert\psi(\tau)\rVert}$, with instantaneous expectation values $\smash{\langle\hat A\rangle\equiv\bra{\phi(\tau)}\hat A\ket{\phi(\tau)}}$, the resulting dynamics is written as
\begin{equation}\label{eq:evo}
    \partial_\tau\ket{\phi(\tau)}=-\hat G\ket{\phi(\tau)}\;,
\end{equation}
where the state-dependent Hermitian generator $\hat G=\hat G^\dagger$ satisfies $\langle\hat G\rangle=0$. For (normalized) ITE, Eq.~\eqref{eq:evo} yields $\smash{\hat G_{\rm ITE}=\bar H}$, with $\smash{\bar H=\hat H-E(\tau)}$ and $\smash{E(\tau)\equiv\langle\hat H\rangle}$. The energy decays at a rate determined solely by its variance $\smash{\dot E_{\rm ITE}=-2\mu_2}$, where $\smash{\mu_m\equiv\langle\bar H^m\rangle}$ denotes the $m$th central moment of the instantaneous energy distribution. 

% ---------------------------------------
% 3) AITE and ITE
% ---------------------------------------

Recently, Benavides-Riveros \etal introduced augmented imaginary-time evolution (AITE)~\cite{benavides_riveros_augmentingimaginarytimeevolution_2026}, which exploits higher moments of the energy distribution through a quadratic, state-dependent generator
\begin{equation}
    \hat G_{\rm AITE} = a\bar H+b(\bar H^2-\mu_2)\;,
    \label{eq:aite_generator}
\end{equation}
with coefficients $\smash{a=\cos(2\varphi)}$ and $\smash{b=\sin(2\varphi)/(2\sqrt{\mu_2})}$, where $\smash{\tan(2\varphi)=\kappa/2}$ and $\smash{\kappa=\mu_3/\mu_2^{3/2}}$ is the Fisher-Pearson skewness coefficient. The resulting local energy flow is a generalization of the double-bracket formulation~\cite{xiaoyue_strategiesoptimizingdoublebracket_2024,gluza_doublebracketquantumalgorithms_2026} and yields
% \cite{Zander2026}
%
\begin{equation}
    \dot E_{\mathrm{AITE}} = -2\mu_2\sqrt{1+\kappa^2/4}
    \leq
    \dot E_{\mathrm{ITE}}\;,
    \label{eq:aite_rate}
\end{equation}
with strict inequality whenever $\kappa \ne 0$. We quantify convergence by the absolute energy error $\epsilon = E - E_0$ and the relative energy error
$\epsilon_{\mathrm{rel}} = |\epsilon|/|E_0|$, where $E_0$ denotes the ground-state energy. Remarkably, unlike ITE, which converges only asymptotically in gapped systems, AITE achieves $\epsilon = 0$ at a finite imaginary time $\tau_*$ once the energy drops below the first excited-state energy~\cite{benavides_riveros_augmentingimaginarytimeevolution_2026}.

Can AITE's accelerated convergence survive variational NQS projection, finite sampling, and finite-step discretization? Conventional NQS optimization relies on the \expl (forward) Euler steps, which evaluate the time derivative at the present state and are subject to stability constraints set by the spectral bandwidth $W_G$ of $\smash{\hat G}$. For AITE, this bandwidth diverges as $\epsilon \to 0$. To overcome this limitation, we project \impl (backward) Euler steps, which instead evaluate the time derivative at the updated state, onto the NQS manifold and compare the resulting \impl ITE (iITE) and \impl AITE (iAITE) schemes. Both \impl schemes substantially accelerate convergence, with iAITE showing consistent gains over iITE as the ansatz becomes more expressive, demonstrating that AITE's acceleration can survive variational projection. 
% The iITE update is closely related to Rayleigh--Gauss--Newton optimization~\cite{webber_rayleighgaussnewtonoptimizationenhanced_2022}, and the projected inverse iteration~\cite{zhang_projectedinverseiteration_2026a,medvidovic_adiabatictransportneural_2026}, which approach ground-state optimization through variational energy minimization and projected eigenvalue iteration, respectively.

% ---------------------------------------
% (5) Models
% ---------------------------------------
\noindent\textbf{\textit{Model Hamiltonians.}} In what follows, we study the square lattice (i) transverse-field Ising (TFI) model~\cite{carleo_solvingquantummanybody_2017}, and (ii) $J_1$-$J_2$ Heisenberg antiferromagnet~\cite{choo_studytwodimensionalfrustrated_2019}
% westerhout_neuralquantumstates_2020, rende_foundationneuralnetworksquantum_2025a}
at $N=L_x\times L_y$ sites. In both cases, we use the $\sigma^z$ product basis, with configurations $\mathbf{s}=(s_1,\ldots,s_N);\; s_\ell=\pm 1$, and impose periodic boundary conditions (PBC). In addition, for the $J_1$-$J_2$ model we apply the Marshall sign rotation~\cite{marshall_antiferromagnetism_1955}. We define $\hat{\boldsymbol S}_\ell=\hat{\boldsymbol\sigma}_\ell/2$, where $\hat{\boldsymbol\sigma}_\ell=(\hat\sigma_\ell^x,\hat\sigma_\ell^y,\hat\sigma_\ell^z)$. Details on lattices, Hamiltonians, symmetry sectors, and variational ans\"atze are given in the \emph{Supplemental Material} (SM)~\cite{SM}.

\begin{figure}[!t]
    \centering
    \includegraphics[width=0.97\columnwidth]{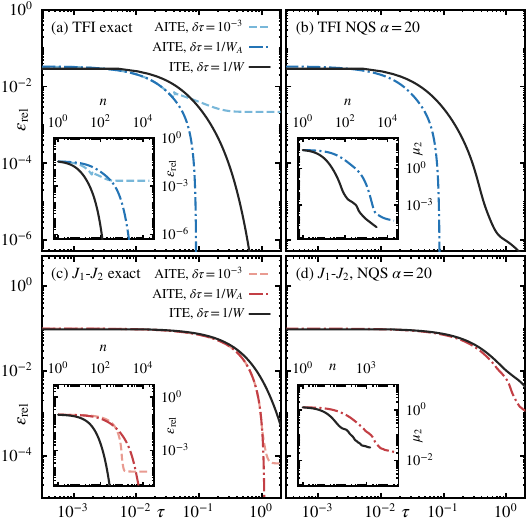}
    \caption{Relative energy error $\epsilon_{\rm rel}$ for finite-step \expl AITE and ITE of the square lattice $L_x=L_y=4$ (a),(b) TFI model at $h=4.0$ and (c),(d) $J_1$-$J_2$ model at $J_2=0.5$ as a function of imaginary time $\tau$. (a),(c) Exact evaluation of \expl [Eq.~\eqref{eq:forward_euler}] ITE at $\delta \tau = 1/W$ (solid lines), AITE at fixed $\delta \tau = 10^{-3}$ (dashed lines), and capped $\delta \tau = 1/W_A$ (dash-dotted lines) time steps for TFI and $J_1$-$J_2$ models, respectively. (b),(d) Same for variational evolution [Eq.~\eqref{eq:sr}] using symmetry-adapted RBM ($\alpha=20$) for both models, respectively, with AITE shown at $\delta\tau=1/W_A$ only. \emph{Full summation} is used. \textit{Insets}: (a),(c) $\epsilon_{\rm rel}$, and (b),(d) Energy variance $\mu_2$ as a function of discrete time steps $n$.
    }
    \label{fig1}
\end{figure}

\begin{figure}[t]
    \centering
    \includegraphics[width=0.96\columnwidth]{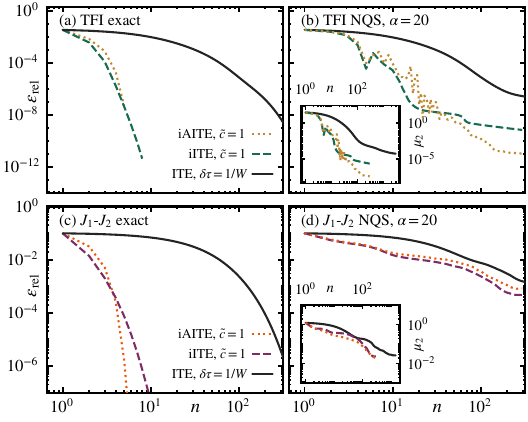}
    \caption{Relative energy error $\epsilon_{\rm rel}$ for finite-step iITE, iAITE, and \expl ITE for the square lattice $L_x=L_y=4$ (a),(b) TFI model at $h=4.0$ and (c),(d) $J_1$-$J_2$ model at $J_2=0.5$ as a function of discrete time steps $n$. (a),(c) \textit{Exact} \expl ITE [Eq.~\eqref{eq:forward_euler}] at $\delta \tau = 1/W$ (solid lines), and \textit{exact} \impl [Eq.~\eqref{eq:backward_euler}] ITE (dashed lines) and AITE (dotted lines) at $\delta\tau_0=1$ with the variance-capped step $\delta\tau_n=\min(\delta\tau_0,\tilde c/\sqrt{2\operatorname{Var}(G_n)})$ and $\tilde c=1$. (b),(d) Same for symmetry-adapted RBM ($\alpha=20$). \textit{Full summation} is used. \textit{Insets}: Energy variance $\mu_2$ as a function of $n$.
    }
    \label{fig2}
\end{figure}

\begin{figure}[t]
    \centering
    \includegraphics[width=\columnwidth]{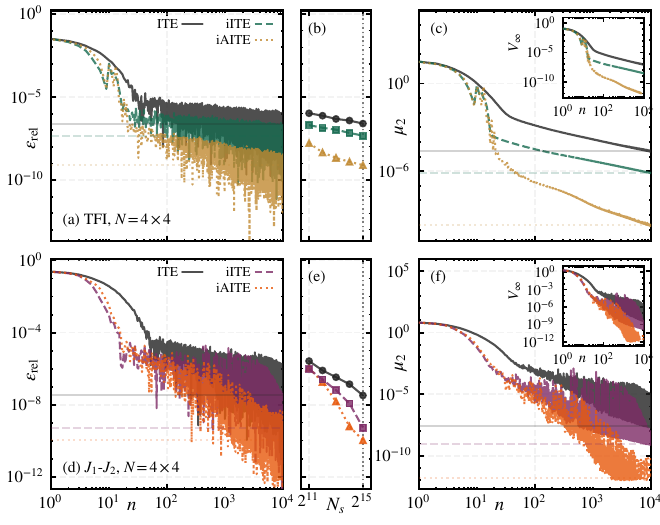}
    \caption{Variational imaginary-time evolution versus discrete steps $n$ for ViT~\cite{viteritti_transformervariationalwave_2023} using \expl ITE (solid lines), iITE (dashed lines), and iAITE (dotted lines) with $N_s$ MCMC samples. Relative energy error $\epsilon_{\rm rel}$ and energy variance $\mu_2$ using the square-lattice $L_x=L_y=4$ (a)--(c) TFI model at $h=4.0$ and (d)--(f) $J_1$-$J_2$ model at $J_2=0.2$. (a),(d) Relative energy error versus discrete steps $n$ for $N_s=32768$ samples. (b),(e) Average $\bar \epsilon _{\rm rel}$ using $5\%$ last steps versus $N_s$. (c),(f) Energy variance $\mu_2$. \textit{Insets}: V-score  $V_{\infty}$~\cite{wu_variationalbenchmarks_2024}.
    }
    \label{fig3}
\end{figure}

\begin{figure*}[!ht]
    \centering
    \includegraphics[width=0.95\textwidth]{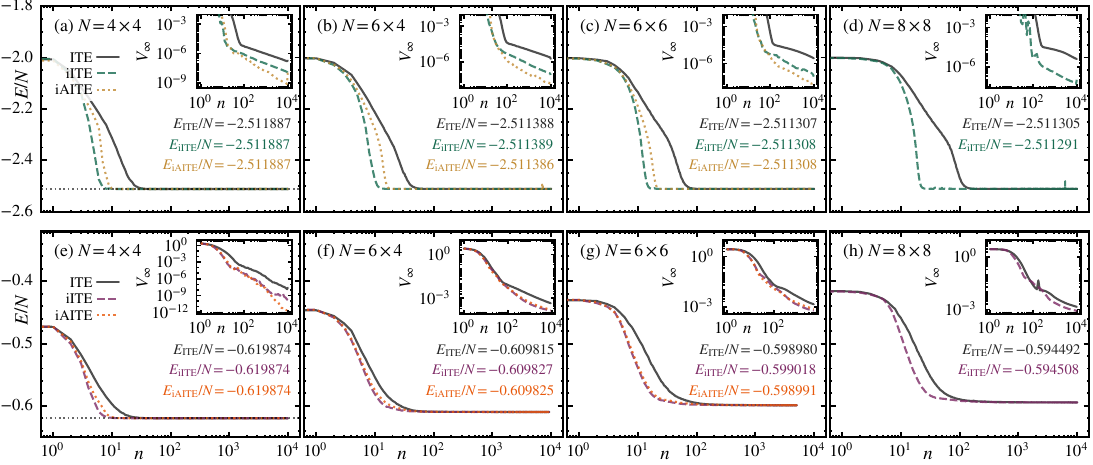}
    \caption{Energy per site $E/N$ as a function of discrete steps $n$ for ViT~\cite{viteritti_transformervariationalwave_2023} using \expl ITE (solid lines), iITE (dashed lines), and iAITE (dotted lines, for $N\le 6\times6$) at different system sizes (a),(e) $N=4\times 4$; (b),(f) $ 6\times 4$; (c),(g) $6\times 6$; (d),(h) $8\times 8$ on a square lattice. Panels (a)--(d) show the results for the TFI model at $h=2.0$ and (e)--(h) for the $J_1$-$J_2$ model at $J_2=0.2$, with the estimated $E_{(\cdots)}/N$ averaged from the last $n_{\rm E}=500$ steps using $N_s=16\,384$ MCMC samples per step, $\delta\tau_0=1$, $c=1.5$ for \expl ITE, and $\tilde c=1$ for iITE and iAITE. Dotted lines show exact $E_0/N$ where available. \textit{Insets}: V-score $\smash{V_{\infty}}$~\cite{wu_variationalbenchmarks_2024}.
    }
    \label{fig4}
\end{figure*}

We first consider the simpler, unfrustrated TFI model,
\begin{equation}
    \hat H_{\mathrm I} = -J\sum_{\langle \ell,\ell'\rangle} \hat\sigma_\ell^z\hat\sigma_{\ell'}^z -h\sum_\ell\hat\sigma_\ell^x\;,
    \label{eq:ising}
\end{equation}
where $\langle\ell,\ell'\rangle$ denotes nearest-neighbor bonds. The model is not exactly solvable and features a quantum critical point at $h_c/J\simeq3.04$ in the thermodynamic limit~\cite{blote_clustermontecarlo_2002}.
% schmitt_quantumphase_2022a}
We fix $J=1$ and consider $h=4 > h_c$ in the paramagnetic phase (Figs.~\ref{fig1}--\ref{fig3}) and $h=2<h_c$ in the ferromagnetic phase (Fig.~\ref{fig4}). To probe a more challenging setting, we select the $J_1$-$J_2$ Heisenberg model~\cite{chandra_possiblespinliquidstate_1988},
\begin{equation}
    \hat H_{J_1J_2} = J_1\sum_{\langle \ell,\ell'\rangle} \hat{\boldsymbol S}_\ell\cdot\hat{\boldsymbol S}_{\ell'} + J_2\sum_{\langle\!\langle \ell,\ell' \rangle\!\rangle} \hat{\boldsymbol S}_\ell\cdot\hat{\boldsymbol S}_{\ell'}\;,
    \label{eq:j1j2}
\end{equation}
where $\langle\!\langle \ell,\ell'\rangle\!\rangle$ denotes next-nearest-neighbor bonds. We fix $J_1=1$ and consider $J_2=0.2$ in the N\'eel-ordered phase (Figs.~\ref{fig3} and~\ref{fig4}) and $J_2=0.5$ in the strongly frustrated regime (Figs.~\ref{fig1} and~\ref{fig2})~\cite{gong_plaquetteorderedphase_2014, choo_studytwodimensionalfrustrated_2019}.
% ============================================================
% FIG. 1: AITE acceleration and explicit finite-step limitation
% ============================================================
%

% ---------------------------------------
% (5) Variational ITE
% ---------------------------------------
\noindent\textbf{\textit{Variational imaginary-time evolution.}} We represent the many-body wavefunction as $\smash{\Psi_{\bm{\theta}}(\mathbf{s})=\braket{\mathbf{s}}{\Psi_{\bm{\theta}}}}$, parametrized by generally complex parameters $\bm{\theta}$, where $\mathbf{s}$ denotes configurations in a chosen computational basis $\smash{\{\ket{\mathbf{s}}\}}$ of $N$ spins~$\tfrac{1}{2}$. The Rayleigh-Ritz variational principle then yields an upper bound to the ground state energy $E_0\leq E_{\mathrm{var}}=\min_{\btheta}\bra{\Psi_{\btheta}}\hat H\ket{\Psi_{\btheta}}/\langle \Psi_{\btheta}|\Psi_{\btheta}\rangle $~\cite{ritz_uebereineneuemethode_1909,macdonald_successiveapproximationsrayleighritz_1933,becca_quantummontecarlo_2017}. In the following, we denote parametrized states using uppercase Greek symbols and lowercase for arbitrary Hilbert-space states.

When restricted to the variational manifold $\mathcal{M}$, the exact ITE direction is projected onto its tangent space using the time-dependent variational principle~\cite{haegeman_timedependentvariationalprinciple_2011,mcardle_variationalansatzbasedquantum_2019a,hackl_geometryvariationalmethods_2020}, resulting in stochastic reconfiguration (SR) for NQS~\cite{sorella_generalizedlanczosalgorithm_2001,sorella_wavefunctionoptimization_2005,carleo_solvingquantummanybody_2017}. SR preconditions the variational force with the quantum geometric tensor (QGT) $S$ and is equivalent to natural-gradient optimization on $\mathcal{M}$~\cite{provost_riemannianstructuremanifolds_1980,amari_naturalgradientworks_1998,stokes_quantumnaturalgradient_2020,gacon_simultaneousperturbationstochastic_2021,fitzek_optimizingvariationalquantum_2024}. Recent developments improved its scalability and conditioning~\cite{chen_empoweringdeepneural_2024,rende_simplelinearalgebra_2024,park_geometrylearningneural_2020} or modified $\mathcal{M}$~\cite{rende_finetuningneuralnetwork_2024,kozic_exploringeffectbasis_2025,cantori_adaptivebasissamplebasedneural_2026}, while keeping the projected updates as an \expl discretization. The link between SR and ITE suggests a complementary, nonexclusive route --- accelerating the underlying evolution itself.

The NQS use NN to parametrize the wavefunction with a large number $N_p$ of variational parameters. The variational state is $\smash{\ket{\Psi_{\btheta}}=\sum_{\mathbf{s}}\Psi_{\btheta}(\mathbf{s})\ket{\mathbf{s}}}$, and VMC estimates the expectation values by sampling from $\smash{p_{\btheta}(\mathbf{s})\propto|\Psi_{\btheta}(\mathbf{s})|^2}$. To isolate the role of imaginary-time dynamics, we first use symmetry-adapted complex restricted Boltzmann machines (RBMs)~\cite{carleo_solvingquantummanybody_2017,nomura_helpingrestrictedboltzmann_2021,choo_symmetriesmanybodyexcited_2018} with hidden-unit density $\alpha$ and parameters shared across $N$ lattice translations. Unless stated otherwise, we use exact Hilbert-space summation for $N\leq16$ to separate variational projection from stochastic effects, and Markov chain Monte Carlo (MCMC) sampling for larger systems. We present RBMs with exact summation and vision transformers (ViTs)~\cite{viteritti_transformervariationalwave_2023} with MCMC sampling. The numerical implementation uses NetKet~\cite{netket3:2022,netket2:2019}, JAX~\cite{jax2018github}, and Flax~\cite{flax2020github}, with our own evolution algorithms.

\noindent\textbf{\textit{Explicit stochastic reconfiguration.}} Projecting Eq.~\eqref{eq:evo} onto the tangent space of the NQS manifold yields the SR equation~\cite{haegeman_timedependentvariationalprinciple_2011,sorella_wavefunctionoptimization_2005}, here written for a holomorphic ansatz. For real-valued parameters such as those of ViT, the real parts of $S$ and $\mathbf F$ are used,
\begin{equation}
    (S+\lambda I)\mathbf d = -\mathbf F[\hat G]\;, 
    \quad S_{kl} = \operatorname{Cov}(O_k^*,O_l)\;,
    \label{eq:sr}
\end{equation}
where $\operatorname{Cov}(X,Y)\equiv\langle XY\rangle - \langle X \rangle\langle Y \rangle$, $\smash{\mathbf d\equiv \partial _\tau {\btheta}}$ represents the variational update direction, $\smash{O_k=\partial_{\theta_k}\log\Psi_{\btheta}}$ yields the logarithmic derivatives, $S$ is the QGT~\cite{provost_riemannianstructuremanifolds_1980}, and $\smash{\lambda\in \mathbb{R}_+}$ serves to regularize the linear system~\cite{sorella_wavefunctionoptimization_2005,schmitt_quantummanybodydynamics_2020}. The variational force is $\smash{F_j[\hat G]=\operatorname{Cov}(O_j^*,G_{\rm loc})}$, where $\smash{G_{\rm loc}}$ is the local estimator of $\smash{\hat G}$. 

At fixed $\btheta$, the matrix $S$ does not depend on the generator, so \expl ITE and AITE differ only through variational forces $\mathbf{F}$. For AITE, this yields
\begin{equation}
    \mathbf F_{\mathrm{AITE}} = a\,\mathbf F[\hat H] +
    b\left[\mathbf F[\hat H^2] - 2E\,\mathbf F[\hat H] \right]\;.
    \label{eq:aite_force}
\end{equation}
The moments entering $a$ and $b$ are estimated from the same Monte Carlo samples, while the quadratic contribution generally requires evaluating $\hat H^2$. The local estimators are given in the SM~\cite{SM}.
% ============================================================
% FIG. 2: implicit schemes -- exact Hilbert space
% ============================================================
%
% ---------------------------------------
% (6) Finite time steps
% ---------------------------------------

\noindent\textbf{\textit{Finite time steps.}} The continuous-time SR equation follows from McLachlan's variational principle, which minimizes the norm of the residual of Eq.~\eqref{eq:evo}, projecting it onto the variational tangent space~\cite{mclachlan_variationalsolutiontimedependent_1964,haegeman_timedependentvariationalprinciple_2011}. Although alternative variational principles and projection strategies exist~\cite{yuan_theoryvariationalquantum_2019}, we focus on finite-step discretizations (see End Matter), which produce discrete SR after introducing the imaginary-time grid $\tau_{n+1}=\tau_n+\delta\tau$, with $\tau_0=0$ and $n=0,1,2,\ldots$, where the step $\delta\tau$ may change with $n$. For the normalized state $\ket{\phi_n}\equiv\ket{\phi(\tau_n)}$, the \expl Euler approximation then gives~\cite{sorella_greenfunctionmonte_1998,ledinauskas_scalableimaginarytime_2023},
%],
\begin{equation}
    \ket{\phi_{n+1}} \simeq \bigl(1-\delta\tau\hat G_n\bigr)\ket{\phi_n}\;, \quad \hat G_n\equiv\hat G[\phi_n]\;,
    \label{eq:forward_euler}
\end{equation}
where both the generator and update direction are evaluated at present $\ket{\phi_n}$. Projecting this discretized evolution onto the variational tangent space yields \expl SR update $\btheta_{n+1}=\btheta_n+\delta\tau\mathbf d_n$, with $\mathbf d_n$ determined from Eq.~\eqref{eq:sr}.

For stiff problems, the \expl Euler method is only conditionally stable~\cite{hairer_solvingordinarydifferential_1991}. Namely, denoting the spectral bandwidth of $\hat G_n$ by $W_G$, the characteristic asymptotic \expl Euler stability restriction, a property of the discreteness rather than of the physics, is $\delta\tau<2/W_G$ near the ground state (see End Matter). For ITE, $\smash{W_G\rightarrow W\equiv E_{\max}-E_0}$ is the bandwidth of the Hamiltonian. In contrast, for AITE, $W_G \rightarrow W_A$ changes with $n$ and diverges as the ground state is approached, $\epsilon \rightarrow 0$,
\begin{equation}
    W_A\sim\frac{W^2}{2\sqrt{\Delta_*\epsilon}}\;,
    \quad
    \delta\tau_{\max}\sim\frac{4\sqrt{\Delta_*\epsilon}}{W^2}\;,
    \label{eq:aite_stability}
\end{equation}
where $\Delta_*=\smash{E_*-E_0}$ is the accessible gap, \ie the gap to the lowest excitation with nonzero overlap with the state (see SM~\cite{SM}), and $W_A$ links the low-energy scale $\Delta_*$ with the total bandwidth $W$. Consequently, the AITE flow becomes increasingly stiff, with the typical stable \expl step vanishing as $\smash{\delta\tau_{\max}\propto\sqrt{\epsilon}}$. At its limit, \expl AITE thus advances less per step than \expl ITE, by a factor $\Delta_*/W$ (see End Matter). This motivates the \impl scheme introduced below, which maintains finite step sizes near convergence.

To illustrate the dynamics and stiffness induced by different generators $\smash{\hat G_n}$, Fig.~\ref{fig1} compares \expl ITE and AITE for both models, for exact Hilbert-space evolution [Figs.~\ref{fig1}(a),(c)] and symmetry-adapted RBM [Eq.~\eqref{eq:sr}] with \textit{full summation} [Figs.~\ref{fig1}(b),(d)]. With the adapted step $\delta\tau=1/W_A$, exact AITE converges faster than ITE in imaginary time $\tau$, consistently with Eq.~\eqref{eq:aite_rate}~\cite{benavides_riveros_augmentingimaginarytimeevolution_2026}. The step, however, shrinks with $W_A$, and AITE may take substantially more iterations $n$ than ITE to reach a given $\epsilon _{\rm rel}$ [insets of Figs.~\ref{fig1}(a),(c)]. At fixed $\delta\tau=10^{-3}$, the growing $W_A$ eventually violates the stability bound [Eq.~\eqref{eq:aite_stability}], leading to overshooting of the excited-state amplitude, so that $\epsilon_{\rm rel}$ stalls at a finite value (see SM~\cite{SM}). In all runs, $W$ and $W_A$ in step sizes are replaced by computable upper bounds (see SM~\cite{SM}), and variational projection adds representability constraints that further limit convergence [Figs.~\ref{fig1}(b),(d)].

% ============================================================
% FIG. 3: Stochastic NQS and size scaling
% ============================================================

\noindent\textbf{\textit{Implicit stochastic reconfiguration.}} A common approach to overcoming the conditional stability of \expl ODE solvers is to use \impl time stepping~\cite{hairer_solvingordinarydifferential_1991,butcher_numericalmethodsordinary_2003,leong_variationalquantumevolution_2022}. We derive it by replacing the \expl Euler update in Eq.~\eqref{eq:forward_euler} with its \impl counterpart. Unlike the \expl Euler scheme, this method evaluates the time derivative at the \emph{future} state, $\ket{\phi _{n+1}} - \ket{\phi _n} \simeq -\delta \tau \hat G_{n+1} \ket{\phi _{n+1}}$, which implies
\begin{equation}
    \bigl(1+\delta\tau\hat G_{n+1}\bigr)\ket{\phi_{n+1}} = \ket{\phi_n}\;.
    \label{eq:backward_euler}
\end{equation}
To avoid a nonlinear \impl solve associated with the state-dependent generator, we approximate it by fixing $\hat G_{n+1} \simeq \hat G _n$,
which acts on the updated state $\ket{\phi_{n+1}}$, while its coefficients are still taken from $\ket{\phi_n}$. We refer to this linearly implicit (backward) Euler step~\cite{hairer_solvingordinarydifferential_1991} as \impl throughout. By expanding $\ket{\phi_{n+1}}$ to the first order in the tangent space about $\ket{\phi_n}$ and then projecting the resulting relation, one obtains the \impl SR equation,
\begin{equation}
    \left(S+\lambda I+\delta\tau K\right)\mathbf d = -\mathbf F\;, \quad K_{jk}=\langle T_j|\hat G_n|T_k\rangle\;,
    \label{eq:implicit_sr}
\end{equation}
where $\ket{T_k}$ denotes the tangent vector orthogonal to the current state; see SM~\cite{SM} for numerical details.
% The regularization and numerical solution of the sampled system, using dense factorizations or, for large systems, the Arnoldi process, are specified in the SM~\cite{SM}.
Unlike Eq.~\eqref{eq:sr}, recovered as $\delta\tau\to0$, Eq.~\eqref{eq:implicit_sr} requires the generator on both sides of the equation, through the variational force $\mathbf F$ and tangent-space matrix $K$. We retain a local approximation at $\tau$, avoiding the nonlinear optimization required by finite-step variational projections based on overlap optimization~\cite{sinibaldi_unbiasingtimedependentvariational_2023}; see End Matter.

While \expl Euler methods require $W_G$ dependent restriction, for a fixed positive-semidefinite generator $\hat G$, Eq.~\eqref{eq:backward_euler} is unconditionally stable. For centered, state-dependent $\smash{\hat{G}}$, however, this result does not automatically extend to the projected \impl scheme [Eq.~\eqref{eq:implicit_sr}], which fixes the generator, uses a linear tangent-space approximation, and is estimated from samples~\cite{chassagneux_numericalstabilityanalysis_2014}. Nevertheless, in the noiseless two-level limit, the iAITE step satisfies $\epsilon_{n+1}\simeq 4\epsilon_n^2/(\delta\tau^2\Delta_*^3)$ at fixed $\delta\tau>0$, whereas iITE converges linearly, $\epsilon_{n+1}\simeq\epsilon_n/(1+\delta\tau\Delta_*)^2$; see SM~\cite{SM}. The divergence of $W_A$ that forces $\delta\tau \to 0$ in the \expl scheme is the origin of the quadratic convergence of the \impl one, since the stiff eigenvalue enters the denominator of the \impl factor $\smash{(1+\delta\tau\hat G_n)^{-1}}$. The overhead of constructing $K$ to solve the linear system in Eq.~\eqref{eq:implicit_sr}, and the increased sampling requirements remain below a factor of two for iITE, but can be greater for iAITE due to the evaluation of $\hat H^2$. For ITE, Eq.~\eqref{eq:implicit_sr} is related to Rayleigh--Gauss--Newton optimization~\cite{webber_rayleighgaussnewtonoptimizationenhanced_2022}, with the penalty parameter identified as $1/\delta\tau$, and also to projected inverse iteration~\cite{zhang_projectedinverseiteration_2026a,medvidovic_adiabatictransportneural_2026}, which came to our attention when developing this approach, with shift $s_n=E_n-1/\delta\tau$, up to regularization. Our imaginary-time formulation, at the same time, extends this to other generators $\smash{\hat G}$, including AITE's $\smash{\hat G _A}$, and uses a variance-based $\delta \tau$; see SM~\cite{SM}.

In Fig.~\ref{fig2}, we illustrate how the \impl schemes remove the limitation of the \expl schemes, where both iITE and iAITE use $\delta\tau_0=1$ and $\tilde c=1$ in the variance-based step cap (see SM~\cite{SM}), while \expl ITE uses $\delta\tau=1/W$. For \textit{exact} evolution [Figs.~\ref{fig2}(a),(c)], both reach a given $\epsilon_{\rm rel}$ in two orders of magnitude fewer steps $n$ than \expl ITE. Projection onto the symmetry-adapted RBM manifold ($\alpha=20$) with \textit{full summation} limits the attainable accuracy but preserves the gain over \expl ITE in both models [Figs.~\ref{fig2}(b),(d)]. The energy variance $\mu_2$ (insets) is systematically lower for the \impl schemes. 
% ============================================================
% FIG. 4: Stochastic NQS - different sizes
% ============================================================

The advantage survives stochastic sampling (Fig.~\ref{fig3}), where MCMC sampling with ViT~\cite{viteritti_transformervariationalwave_2023} replaces full summation, for the TFI model at $h=4$ [Fig.~\ref{fig3}(a)--(c)] and the $J_1$-$J_2$ model at $J_2=0.2$ [Fig.~\ref{fig3}(d)--(f)]. At $N_s=2^{15}$ samples [Figs.~\ref{fig3}(a),(d)], both \impl schemes outperform \expl ITE, with iAITE ultimately achieving the lowest $\epsilon _{\rm rel}$ for the TFI model [Fig.~\ref{fig3}(a)--(c)]. This advantage persists across sampling budgets $N_s$ [Figs.~\ref{fig3}(b),(e)], although statistical fluctuations and limited ansatz capacity again constrain the attainable accuracy, particularly for the more challenging $J_1$-$J_2$ model [Figs.~\ref{fig3}(d)--(f)]. For iAITE, sampling noise enters through the moments $\mu_2$ and $\mu_3$ that set $a$ and $b$, through the estimator of $\hat H^2$, and through the estimate of $K$, which is generally not Hermitian at finite $N_s$ (see SM~\cite{SM}). The energy variance [Figs.~\ref{fig3}(c),(f)] and corresponding V-scores shown in the insets, $\smash{V_{\infty}=N\mu_2/(E-E_\infty)^2}$, where $E_\infty=\operatorname{Tr}(H)/\dim\mathcal H$ is the infinite-temperature energy of the simulated Hilbert space (the $S^z=0$ sector for $J_1$-$J_2$)~\cite{wu_variationalbenchmarks_2024}, further characterize the quality of the converged variational states. 

Finally, Fig.~\ref{fig4} extends the comparison from $4\times4$ to $8\times8$ for the TFI model at $h=2$ [Figs.~\ref{fig4}(a)--(d)] and the $J_1$-$J_2$ model at $J_2=0.2$ [Figs.~\ref{fig4}(e)--(h)], with iAITE shown up to $6\times6$. At each size, both \impl schemes reach the plateau of $E/N$ in several times fewer steps $n$ than \expl ITE, and the V-score drops earlier and further (insets). On larger lattices, the energies obtained with different schemes agree within $\sim 10^{-5}$ per site, with iAITE trajectories closely following those of iITE. The convergence gains of \impl schemes persist with increasing $N$, while greater ansatz expressivity shall yield further improvements, particularly for iAITE.
% ============================================================================
% (9) Discussion
% ============================================================================

\textbf{\textit{Discussion.}} \Expl Euler steps cannot use the faster flow of AITE, because its stiffness forces $\delta \tau \to 0$ as $\epsilon \to 0$. We combine \impl Euler steps with Galerkin projection~\cite{koch_dynamicallowrankapproximation_2007,lubich_quantumclassicalmolecular_2008, kieri_discretizeddynamicallowrank_2016} to remove this limitation, yielding an \impl SR scheme for general Hermitian $\smash{\hat G}$ driven flows [Eq.~\eqref{eq:evo}]. We demonstrate that the acceleration of AITE~\cite{benavides_riveros_augmentingimaginarytimeevolution_2026} over ITE largely survives finite time steps, variational projection, and MCMC sampling in the tested models. In \textit{exact} evolution, both \impl schemes reach a given $\epsilon _{\rm rel}$ in about two orders of magnitude fewer steps than \expl ITE, and they preserve the gain over \expl ITE for projected NQS with \textit{full summation}. These gains persist under MCMC sampling with finite $N_s$, although acceleration requires an ansatz $\Psi _{\btheta}$ whose tangent vectors can accurately capture the direction of the flow generated by $\smash{\hat G}$. 
% Only then do the \impl schemes consistently improve convergence.

While iAITE can outperform iITE, evaluating $\hat H^2$ and higher energy moments $\mu_2$ and $\mu_3$ requires additional computational effort and careful stochastic estimation. Improved importance sampling~\cite{trail_heavytailedrandomerror_2008,misery_lookingelsewhereimproving_2025} may help partially mitigate this in the future. Meanwhile, iITE retains a modest computational overhead relative to conventional SR, consistent with related RGN results~\cite{webber_rayleighgaussnewtonoptimizationenhanced_2022}.

Our results establish \impl variational imaginary-time evolution as a general and promising route to improved ground-state optimization. Advancements in sampling, linear solvers, and variational ansätze $\Psi _{\btheta}$ may extend its reach to much larger and more challenging systems, including small-gap regimes~\cite{zhang_projectedinverseiteration_2026a}. Promising future directions include fermionic~\cite{choo_fermionicneuralnetwork_2020a,pfau_abinitiosolutionmanyelectron_2020} and non-Hermitian~\cite{solinas_biorthogonalneuralnetwork_2026} systems in condensed matter physics and quantum chemistry, such as transcorrelated Hamiltonians of solids and lattice models~\cite{tsuneyuki_transcorrelatedmethodanother_2008,dobrautz_compactnumericalsolutions_2019,baiardi_transcorrelateddensitymatrix_2020,sokolov_ordersmagnitudeincreased_2023} and of molecules~\cite{boys_determinationenergieswavefunctions_1969,cohen_similaritytransformationelectronic_2019,motta_quantumsimulationelectronic_2020,haupt_optimizingjastrowfactors_2023,christlmaier_xtcefficienttreatment_2023,dobrautz_realchemicalaccuracy_2024,magnusson_efficientquantumcomputing_2024,liao_interpolativeseparabledensityfitting_2026,hong_wavefunctionregularityeigenvector_2026}.\\

\textbf{\textit{Data and code availability.}} The simulation code and the scripts used to generate the figures~\cite{zenodo} will be made publicly available upon publication, while trained network parameters and larger numerical datasets are available upon reasonable request.

\textbf{\textit{Use of AI-assisted tools.}} AI tools, including Claude Opus 5.5 (Anthropic) and ChatGPT (OpenAI), were used to assist with English-language polishing, numerical test development, and optimization of linear solvers. All AI-assisted contributions were reviewed and validated by the authors.

%============================================================================

\begin{acknowledgments} 

\textbf{\textit{Acknowledgments.}} 
We acknowledge funding from the German Federal Ministry of Research, Technology and Space (BMFTR) under the research program Quantensysteme and funding measure Quantum Futur 3 for project No. 13N17229 as well as the Helmholtz Association via the Initiative and Networking Fund for projects No. VH-NG-21-08 under the Helmholtz Investigator program as well Projects KA-QUS-02 (qFLOW) and KA-QUS-03 (QT-Batt) under the Helmholtz Quantum Use case call.
We  acknowledge the computing time made available to us on the high-performance computers at the NHR Center Paderborn Center for Parallel Computing (PC2) (projects hpc-prf-qehpc and hpc-prf-nniap) and on the Rosi high-performance computing cluster at the Helmholtz-Zentrum Dresden-Rossendorf (HZDR).
PC2 is jointly supported by the Federal Ministry of Research, Technology and Space of Germany and the state governments participating in the NHR (www.nhr-verein.de/unsere-partner). 

\end{acknowledgments}

% =========================
% End Matter (placed before the references for readability)
% =========================

\appendix
\setcounter{secnumdepth}{2}
\renewcommand{\thesection}{\Alph{section}}
\renewcommand{\appendixname}{}

\onecolumngrid
\vspace{0.5em}

\begin{center}
    \rule{0.25\textwidth}{0.4pt}\\[0.8em]
    {\large\bfseries End Matter}
\end{center}

\vspace{1.5em}
\twocolumngrid

%%%%%%%%%%%%%%%
\section{Variational projection and explicit SR}
\label{em:sr}

We obtain the \expl SR update from an \expl Euler step [Eq.~\eqref{eq:forward_euler}] on the time grid $\tau_n$. For the normalized variational state at step $n$, $\ket{\Phi_n} \equiv \ket{\Phi(\btheta_n)} = \ket{\Psi_n}/\lVert\Psi_n\rVert$, the Euler target $\ket{\phi_{n+1}}$ generally falls outside the manifold $\mathcal{M}$~\cite{haegeman_timedependentvariationalprinciple_2011}. At the same time, a parameter shift $\delta\btheta = (\delta \theta _1, \dots, \delta \theta _{N_p})$ gives the first-order expansion of the unnormalized variational state
\begin{equation}
    \ket{\Psi_{n+1}} \simeq \ket{\Psi_n} + \sum_k \delta\theta_k\,\ket{\partial_k\Psi_n}\;,
    \label{eq:em:variational_state}
\end{equation}
where $\ket{\partial_k\Psi}\equiv\partial_{\theta_k}\ket{\Psi}$ denotes the partial derivatives evaluated at step $n$. With variational velocity $\left.\partial_\tau\btheta\right|_{\tau_n}$ and $\delta\btheta=\delta\tau\mathbf d_n$, Eq.~\eqref{eq:em:variational_state} becomes
\begin{equation}
    \ket{\Psi_{n+1}} \simeq \ket{\Psi_n} + \delta\tau\sum_k d_{n,k} \ket{\partial_k\Psi_n}\;.
\end{equation}
%
% MK: Is it true with the global phase - McLachlan in principle concerns the issues with the global phase at real-time evolution
The derivatives carry components parallel to $\ket{\Phi_n}$ that affect only the normalization and global phase~\cite{kramer_geometrytimedependentvariational_1981, hackl_geometryvariationalmethods_2020}. Removing these with $\smash{\Pi_\perp^{(n)} = I - \ket{\Phi_n}\bra{\Phi_n}}$ and accounting for normalization yields the tangent vectors
\begin{equation}
    \begin{aligned}
        \vert {T_{k}^{(n)}} \rangle & \equiv \lVert\Psi_n\rVert^{-1} \Pi _\perp ^{(n)} \ket{\partial_k\Psi_n}  \\ 
        &= \lVert\Psi_n\rVert^{-1}\left (\ket{\partial_k\Psi_n} - \ket{\Phi_n}\langle \Phi_n|\partial_k \Psi_n \rangle\right)\;.
    \end{aligned}
    \label{eq:em:projective_tangent}
\end{equation}
The logarithmic derivatives $\smash{O_k^{(n)}(\mathbf{s})=\partial_{\theta_k}\log\Psi_{\btheta_n}(\mathbf{s})}$ define diagonal operators $\smash{\hat O_k^{(n)}\ket{\mathbf{s}}\equiv O_k^{(n)}(\mathbf{s})\ket{\mathbf{s}}}$, such that $\smash{\ket{\partial_k\Psi_n} = \hat O_{k}^{(n)}\ket{\Psi_n}}$ for both complex and real parameters, and Eq.~\eqref{eq:em:projective_tangent} becomes
% Under a holomorphic ansatz, the logarithmic derivatives $\smash{O_k^{(n)}(\mathbf{s})=\partial_{\theta_k}\log\Psi_{\btheta_n}(\mathbf{s})}$ define diagonal operators $\smash{\hat O_k^{(n)}\ket{\mathbf{s}}\equiv O_k^{(n)}(\mathbf{s})\ket{\mathbf{s}}}$, such that $\smash{\ket{\partial_k\Psi_n} = \sum_{\mathbf{s}} \partial_{\theta_k} \Psi_{\btheta_n}(\mathbf{s})\ket{\mathbf{s}} = \hat O_{k}^{(n)}\ket{\Psi_n}}$, and Eq.~\eqref{eq:em:projective_tangent} becomes
%
\begin{equation}
    |T_k^{(n)}\rangle = \left(\hat O_k^{(n)}-\left\langle O_k^{(n)}\right\rangle\right)\ket{\Phi _n}\;.
    \label{eq:em:tangent_final}
\end{equation}
According to McLachlan's variational principle~\cite{mclachlan_variationalsolutiontimedependent_1964,haegeman_timedependentvariationalprinciple_2011}, $\mathbf d_n$ minimizes the squared residual norm, where $\smash{\vert R^{(n)}\rangle = \sum_k d_{n,k}\vert T_{k}^{(n)}\rangle + \hat G_n\ket{\Phi_n}}$, measuring the deviation from the imaginary-time flow,
\begin{equation}
    \mathcal{R}(\mathbf d_n) \equiv \lVert R^{(n)}\rVert ^2 = \left\lVert \sum_k d_{n,k}|T_k^{(n)}\rangle
    +\hat G_n\ket{\Phi_n} \right\rVert ^2 \;.
    \label{eq:em:residual}
\end{equation}
We then define the variational forces
\begin{equation}
    F_{k}^{(n)}\equiv\langle T_k^{(n)}|\hat G_n\ket{\Phi_n} = \operatorname{Cov} (O_k^{(n)*}, G_{\rm loc}^{(n)})\;,
    \label{eq:em:forces}
\end{equation}
%
%
% \begin{equation}
%     G_{\mathrm{loc}}^{(n)}(\mathbf s)=\langle\mathbf s|\hat G_n|\Phi_n\rangle/\langle\mathbf s|\Phi_n\rangle\;,
%     \label{eq:em:gloc}
% \end{equation}
%
which, together with the local estimator $G_{\mathrm{loc}}^{(n)}(\mathbf s)=\langle\mathbf s|\hat G_n|\Phi_n\rangle/\langle\mathbf s|\Phi_n\rangle$ and QGT $S^{(n)}_{kl} \equiv \langle T^{(n)}_{k}|T^{(n)}_l\rangle = \operatorname{Cov} (O_k^{(n)*}, O_l^{(n)})$, lead to the squared residual, which expands as
\begin{equation}
    \begin{aligned}
        \mathcal{R}(\mathbf d_n)
        & = \sum_{k,l} d_{n,k}^*\,d_{n,l}\,S_{kl}^{(n)} \\
        & + \sum_k\left(d_{n,k}^*F_{k}^{(n)} + d_{n,k}F_{k}^{(n)*}\right) + \langle \hat G_n^2\rangle \;.
    \end{aligned}
\end{equation}
Assuming $d_{n,k}$ and $d_{n,k}^*$ are independent and enforcing stationarity, $\partial _{d_{n,k}^*} \mathcal{R} = 0$, with respect to the latter gives
\begin{equation}
    \sum_l S_{kl}^{(n)}\,d_{n,l} + F_{k}^{(n)} = 0\;,
    \label{eq:em:from_least_squared}
\end{equation}
which recovers the unregularized SR relation $S^{(n)}\mathbf d_n = -\mathbf F^{(n)}[\hat G_n]$. In practice, a diagonal shift $\lambda I$ is introduced to stabilize the solution when $S^{(n)}$ is singular or poorly conditioned, leading to Eq.~\eqref{eq:sr}. When $\btheta$ is restricted to be real, the stationarity condition instead yields $\Re \{S^{(n)}\}\mathbf d_n=-\Re\{\mathbf F^{(n)}[\hat G_n]\}$.

For a holomorphic ansatz, the McLachlan, Dirac-Frenkel, and Galerkin methods coincide on the complex-linear tangent space~\cite{broeckhove_equivalencetimedependentvariational_1988,haegeman_timedependentvariationalprinciple_2011}. In particular, for \expl Euler, the difference between the variational state and the Euler target, divided by $\delta\tau$, is precisely the residual $\ket{R^{(n)}}$ [Eq.~\eqref{eq:em:residual}]. Its Galerkin projection onto the instantaneous tangent space gives~\cite{koch_dynamicallowrankapproximation_2007}
\begin{equation}
        0 = \langle T_j^{(n)}|R^{(n)}\rangle = \sum_k S_{jk}^{(n)}d_{n,k}+F_j^{(n)} \;,
    \label{eq:em:explicit_galerkin}
\end{equation}
which matches the stationarity condition in Eq.~\eqref{eq:em:from_least_squared}.

%%%%%%%%%%%%%%%
\section{Spectral stiffness and different Euler schemes}
\label{em:stiffness}

\paragraph*{\textbf{Explicit Euler scheme.}} We compare the spectral stability of the \expl Euler discretization introduced in Eq.~\eqref{eq:forward_euler} for a Hermitian generator $\hat G_n$. For both ITE and AITE generators, which are functions of $\hat H$, they share its eigenstates, $\hat G_n\ket{E_j} = g_{j,n}\ket{E_j}$. Expanding the normalized state as $\ket{\phi_n}=\sum_j c_{j,n}\ket{E_j}$, where $c_{j,n}=\langle E_j \ket {\phi _n}$, the coefficients evolve as
\begin{equation}
    c_{j,n+1} = \lVert (1-\delta\tau\hat G_n)\ket{\phi_n} \rVert^{-1}(1-\delta\tau g_{j,n}) c_{j,n}\;.
\end{equation}
Normalization cancels in the ratios, leading to
\begin{equation}
    \frac{c_{j,n+1}}{c_{0,n+1}} =\frac{c_{j,n}}{c_{0,n}} \frac{1-\delta\tau g_{j,n}}{1-\delta\tau g_{0,n}} \;,
    \label{em:eq:ratios}
\end{equation}
and the suppression of each excited-state component is damped relative to the ground state iff $\left|(1-\delta\tau g_{j,n})/(1-\delta\tau g_{0,n})\right| < 1$ and thus     $0<\delta\tau < 2 / (g_{j,n}+g_{0,n}),$ for every $j>0$. 
%
% \begin{equation}
%     \left|(1-\delta\tau g_{j,n})/(1-\delta\tau g_{0,n})\right| < 1\;,
%     \label{eq:em:stability_general}
% \end{equation}
% %
Defining $W_G\equiv \max_j g_{j,n} - g_{0,n}$, the centered step remains stable when $0<\delta\tau<2/(W_G+2g_{0,n})$, as long as the denominator is positive. Close to the ground state, $g_{0,n}\to0$, which yields
\begin{equation}
    \delta\tau_{\max}\simeq 2 / W_G \;.
    \label{em:eq:stability}
\end{equation}
For ITE, $W_G=W$, while for AITE, $W_G=W_A$ is set by the instantaneous energy moments and diverges asymptotically, implying that $\delta\tau_{\max} \to 0$.

To estimate the asymptotic time step, we adopt the two-level approximation (see SM~\cite{SM}), where the remaining excited-state population is dominated by the lowest accessible excitation, with the accessible gap $\Delta_*$. As $\epsilon \to 0$, the corresponding skewness diverges as $\kappa\simeq\sqrt{\Delta_*/\epsilon}$, and the optimal AITE parameters satisfy
\begin{equation}
    a\simeq 2\sqrt{\frac{\epsilon}{\Delta_*}}\;,\quad b\simeq \frac{1}{2\sqrt{\Delta_*\epsilon}}\;.
\end{equation}
In this regime, the growing AITE bandwidth becomes
\begin{equation}
    W_A=W[a + b(W - 2\epsilon)] \simeq \frac{W^2}{2\sqrt{\Delta_*\epsilon}}\;,
    \label{em:eq:aite_bandwidth}
\end{equation}
which implies $\delta\tau_{\rm A}^{\max} \simeq 4\sqrt{\Delta_*\epsilon} / W^2$, cf. Eq.~\eqref{eq:aite_stability}. Therefore, even though the continuous-time AITE acceleration diverges as $\mathcal{S}_A=\sqrt{1+\kappa^2/4}\simeq\frac{1}{2}\sqrt{\Delta_*/\epsilon}$ [Eq.~\eqref{eq:aite_rate}], the largest stable \expl Euler step vanishes as $\sqrt{\epsilon}$, and their product remains finite,
\begin{equation} 
    \mathcal{S}_A \delta\tau_{\rm A}^{\max} \simeq 2\Delta_*/W^2\;,
\end{equation}
which is smaller than the ITE value $2/W$ by a factor $\Delta_*/W$.\\

\paragraph*{\textbf{Implicit Euler scheme.}} In contrast, \impl Euler is A-stable and L-stable, so stiff high-energy modes are damped, %\DIFdelbegin \DIFdel{completely}\DIFdelend \DIFaddbegin \DIFadd{in the stiff limit}\DIFaddend , 
unlike in the Crank--Nicolson or implicit-midpoint schemes~\cite{hairer_solvingordinarydifferential_1991}. Close to the ground state, the two generators obey $g_{0,n}\to0^-$ and $g_{j,n}>0$ for $j>0$. In the ITE case, $g_{0,n}=-\epsilon$, while for AITE one finds $g_{0,n}\simeq-\sqrt{\Delta_*\epsilon}/2$. Hence, for any fixed $\delta\tau>0$ and sufficiently close to convergence, we have $0<1+\delta\tau g_{0,n} < 1+\delta\tau g_{j,n}$, which guarantees $|(1+\delta\tau g_{0,n})/(1+\delta\tau g_{j,n})|<1$. As a result, unlike the \expl method, \impl Euler does not require $\delta\tau\to0$ as $\epsilon \rightarrow 0$. In particular, the idealized two-level \impl AITE displays quadratic convergence in $\epsilon$; see SM~\cite{SM}.

%%%%%%%%%%%%%%%
\section{Implicit SR}
\label{em:implicit}

We extend the variational projection to the \impl Euler discretization of Eq.~\eqref{eq:backward_euler}. A fully \impl scheme requires solving self-consistently for both the new state and its state-dependent generator $\hat G_{n+1}=\hat{G}[\Phi_{n+1}]$~\cite{leong_variationalquantumevolution_2022}. Here, we use the linearly implicit step of the main text, fixing the generator at the current state, $\hat G_{n+1}\simeq \hat G_n\equiv \hat G[\Phi_n]$. Freezing the generator does not change the fixed points, since at a fixed point $\Phi_{n+1}=\Phi_n$ and hence $\hat G_{n+1}=\hat G_n$.

Approximating the new state by its first-order tangent-space expansion about $\ket{\Phi_n}$, with normalization and phase fixed at the current $n$'th state, gives
\begin{equation}
    \left(1+\delta\tau\hat G_n\right) \underbrace{\left(\ket{\Phi_n} +\delta\tau\sum_k d_{n,k}|T_k^{(n)}\rangle\right)}_{\simeq \ket{\Phi _{n+1}}} \simeq \ket{\Phi_n}\;.
\end{equation}
Subtracting $\ket{\Phi_n}$ and dividing by $\delta\tau$, one obtains the \impl Euler residual
\begin{equation}
    \vert i^{(n)} \rangle = \hat G_n \ket{\Phi_n} + \sum_k d_{n,k} \left( 1 + \delta\tau\hat G_n\right) |T_k^{(n)}\rangle\;.
    \label{eq:em:implicit_residual}
\end{equation}
Since the tangent space captures all first-order variational displacements, we impose the Galerkin orthogonality~\cite{koch_dynamicallowrankapproximation_2007,lubich_quantumclassicalmolecular_2008, kieri_discretizeddynamicallowrank_2016}, $\langle T_j^{(n)}|i^{(n)}\rangle=0$, for $j=1,\dots,N_p$. This gives

\begin{equation}
    \begin{aligned}
        0  &= \langle T_j^{(n)} |\hat G_n\ket{\Phi_n} + \\
           & \sum_k d_{n,k} \Bigl[\langle T_j^{(n)}|T_k^{(n)}\rangle  + \delta\tau\langle T_j^{(n)} |\hat G_n|T_k^{(n)}\rangle \Bigr] \;.
    \end{aligned}
    \label{eq:em:implicit_sr_discrete}
\end{equation}
which can be expressed equivalently as
\begin{equation}
    \sum_k\left(S_{jk}^{(n)}+\delta\tau K_{jk}^{(n)}\right)d_{n,k} = -F_j^{(n)}\;,
    \label{eq:em:sr_implicit}
\end{equation}
where $\smash{K_{jk}^{(n)}=\langle T_j^{(n)} |\hat G_n|T_k^{(n)}\rangle}$ denotes the tangent-space matrix. Adding diagonal regularization $\lambda I$, one recovers Eq.~\eqref{eq:implicit_sr}. 

When $\btheta$ is real, as for the ViT, the projection of Eq.~\eqref{eq:em:implicit_residual} gives
\begin{equation}
    \Re\left\{S^{(n)}+\delta\tau K^{(n)}\right\}\mathbf d_n = -\Re\left\{\mathbf F^{(n)}\right\}\;,
    \label{eq:em:sr_implicit_real}
\end{equation}
which reduces to $\Re\{S^{(n)}\}\mathbf d_n=-\Re\{\mathbf F^{(n)}\}$ as $\delta\tau\to0$, consistent with the real form of \expl SR.

The projected \impl Euler scheme recovers \expl SR as $\delta\tau \to 0$. Unlike minimization of residual-norms $\smash{\partial _{d_{n,k}^*}\smash{\lVert i ^{(n)}\rVert^2 = 0}}$, the Galerkin projection avoids additional matrix elements involving $\smash{\hat G_n^2}$, reducing the computational and statistical cost. However, the effective matrix $S^{(n)}+\delta\tau K^{(n)}$ need not remain positive definite at large time steps. Residual-norm minimization, regularization, and numerical stability are discussed in the SM~\cite{SM}. 

This approach shares the tangent-space construction of the VMC linear method~\cite{toulouse_optimizationquantummonte_2007,frank_learningneuralnetwork_2021}. Unlike the linear method, which minimizes energy through a generalized eigenvalue problem, our approach projects the \impl Euler equation, yielding a time-step-dependent linear system applicable also to state-dependent generators such as AITE.

% References cited only in the SM, listed in the note of Ref.~[SM]. Kept here so that they come last.
\nocite{pearson_contributionsmathematicaltheory_1895,brockett1991,bloch_completelyintegrablegradient_1992,alghadeer_doublebracketalgorithmiccooling_2025a,richter_violationmarshallpeierlssign_1994,reinhart_goingdeepreaching_2026a,higham_accuracystabilitynumerical_2002,frommer_restartedgmresshifted_1998,marquardt_algorithmleastsquaresestimation_1963}
\bibliography{references}

%%%%%%%%%%%%%%%%%%%%%%%%%%%%%%%%
% SM
%%%%%%%%%%%%%%%%%%%%%%%%%%%%%%%%

\ifdefined\combinedarxiv
    \clearpage
    \setcounter{section}{0}
    \setcounter{subsection}{0}
    \setcounter{equation}{0}
    \setcounter{figure}{0}
    \setcounter{table}{0}
    
    \renewcommand{\thesection}{S\arabic{section}}
    \renewcommand{\theequation}{S\arabic{equation}}
    \renewcommand{\thefigure}{S\arabic{figure}}
    \renewcommand{\thetable}{S\arabic{table}}
    \renewcommand{\thesubsection}{\thesection.\arabic{subsection}}
    
    \onecolumngrid
    
    \begin{center}
        {\Large\bfseries Supplemental Material for: Augmented Imaginary-Time Evolution of Neural Quantum States with Implicit Time Steps}
    \end{center}
    
    \vspace{1em}
    \twocolumngrid

    \makeatletter
    
    \renewcommand{\thesection}{S\arabic{section}}
    \renewcommand{\thesubsection}{\thesection.\arabic{subsection}}
    \renewcommand{\thesubsubsection}{\thesubsection.\arabic{subsubsection}}
    
    \renewcommand{\p@section}{}
    \renewcommand{\p@subsection}{}
    \renewcommand{\p@subsubsection}{}
    
    \makeatother
    \raggedbottom

% ---------------------------------------------------------------------------
% Metadata
% ---------------------------------------------------------------------------
\title{Supplemental Material for:
Augmented Imaginary-Time Evolution of Neural Quantum States
with Implicit Time Steps}

\author{Maksymilian Kliczkowski\orcidlink{0000-0002-7987-9913}}
\affiliation{Center for Advanced Systems Understanding (CASUS), 02826 G\"orlitz, Germany\looseness=-3}
\affiliation{Helmholtz-Zentrum Dresden-Rossendorf (HZDR), 01328 Dresden, Germany\looseness=-3}

\author{Thomas D. K\"uhne}
\affiliation{Center for Advanced Systems Understanding (CASUS), 02826 G\"orlitz, Germany\looseness=-3}
\affiliation{Helmholtz-Zentrum Dresden-Rossendorf (HZDR), 01328 Dresden, Germany\looseness=-3}
\affiliation{Institute of Artificial Intelligence, Technische Universit\"at Dresden, Helmholtzstra{\ss}e 10, 01069 Dresden, Germany\looseness=-3}

\author{Fedor  Šimkovic IV}
\affiliation{IQM Quantum Computers, Georg-Brauchle-Ring 23-25, 80992 Munich, Germany}

\author{Prachi Sharma}
\affiliation{IQM Quantum Computers, Georg-Brauchle-Ring 23-25, 80992 Munich, Germany}

\author{Carlos L. Benavides-Riveros}
\affiliation{IQM Quantum Computers, Georg-Brauchle-Ring 23-25, 80992 Munich, Germany}

\author{Werner Dobrautz\orcidlink{0000-0001-6479-1874}}
\affiliation{Center for Advanced Systems Understanding (CASUS), 02826 G\"orlitz, Germany\looseness=-3}
\affiliation{Helmholtz-Zentrum Dresden-Rossendorf (HZDR), 01328 Dresden, Germany\looseness=-3}
\affiliation{Center for Scalable Data Analytics and Artificial Intelligence (ScaDS.AI), TU Dresden, 01062 Dresden, Germany\looseness=-3}

\date{\today}
\maketitle

% -------------------------------------------
% Supplemental Material
% -------------------------------------------

In what follows, we provide derivations and numerical details supplementing the main text. Sec.~\ref{sm:ite} reviews normalized imaginary-time evolution (ITE), gives convergence bounds, and analyzes two-level asymptotics. Sec.~\ref{sm:aite} details the augmented imaginary-time evolution (AITE)~\cite{benavides_riveros_augmentingimaginarytimeevolution_2026} and Sec.~\ref{sm:hamiltonian_bw} introduces a computable bandwidth (stiffness) bound. Sec.~\ref{sm:nqs} describes the variational ansatz and sampling, Sec.~\ref{sm:implicit_least_squares} justifies the \impl SR equations, Sec.~\ref{sm:time_steps} specifies the Euler time steps, Sec.~\ref{sm:nqs_estimators} introduces the estimators and linear solvers, and Sec.~\ref{sm:two_level_euler} analyzes the local convergence of \impl evolution in the two-level limit. Complementing numerical results are shown in Sec.~\ref{sm:results}.

Unless stated otherwise, unnormalized states are denoted as $\ket{\psi}$, whereas normalized states are $\smash{\ket{\phi}=\ket{\psi}/\sqrt{\braket{\psi}{\psi}}}$. Parametrized states use uppercase Greek symbols, $\{\ket{\Psi}, \ket{\Phi}\}$, and arbitrary states are lowercase. The dependence on the imaginary time $\tau$ is omitted whenever clear from context, \eg $\ket{\phi}\equiv\ket{\phi(\tau)}$ and $E\equiv E(\tau)$.

% -------------------------------------------
% ITE
% -------------------------------------------

\section{Imaginary-time evolution}
\label{sm:ite}

We derive the conventional normalized ITE~\cite{mcardle_variationalansatzbasedquantum_2019a} and its equivalent density-matrix (double-bracket) formulation~\cite{xiaoyue_strategiesoptimizingdoublebracket_2024, Zander2026,gluza_doublebracketquantumalgorithms_2026}. We demonstrate that this flow maximizes the instantaneous energy decrease among the generators $\smash{\hat G}$ whose variance equals that of the Hamiltonian. 

\subsection{Normalized imaginary-time evolution}\label{sm:ite:norm}

Let $\hat{H}=\hat{H}^\dag$ be a time-independent Hermitian Hamiltonian, and assume $\hbar=1$ throughout. Upon analytic continuation to imaginary time~\cite{wick_propertiesbethesalpeterwave_1954,mclachlan_variationalsolutiontimedependent_1964,mcardle_variationalansatzbasedquantum_2019a} $t=-i\tau$, the unnormalized state
\begin{equation}
    \ket{\psi(\tau)} = e^{-\tau\hat{H}}\ket{\psi(0)}\;,
\end{equation}
obeys the imaginary-time Schr\"odinger equation $\partial_\tau\ket{\psi} = -\hat{H}\ket{\psi}$. Its norm $\mathcal{N}(\tau) = \inner{\psi(\tau)}{\psi(\tau)}$ evolves as 
\begin{equation}
    \partial_\tau \mathcal{N} = 
    -2\bra{\psi}\hat{H}\ket{\psi} = -2E(\tau)\mathcal{N}(\tau)\;,
\end{equation}
where $E(\tau)=\expect{\psi(\tau)}{\hat{H}}{\psi(\tau)}/\mathcal{N}(\tau)$ is the instantaneous energy. It follows that $\partial_\tau\sqrt{\mathcal N}=-E(\tau)\sqrt{\mathcal N}$, so the normalized state $\ket{\phi}=\ket{\psi}/\sqrt{\mathcal N}$ obeys
\begin{align}
    \partial_\tau\ket{\phi(\tau)}
    &= \frac{1}{\sqrt{\mathcal{N}}}\,\partial_\tau\ket{\psi(\tau)}
    + \left(\partial_\tau\frac{1}{\sqrt{\mathcal{N}}}\right)\ket{\psi(\tau)}
    \nonumber \\
    &= -\bigl[\hat{H}-E(\tau)\bigr]\ket{\phi(\tau)} \;,
    \label{sm:eq:sm:ite_norm}
\end{align}
and for the centered Hamiltonian $\bar H=\hat H-E$, one has $\expect{\phi}{\bar H}{\phi}=0$, ensuring norm preservation. Differentiating $E=\expect{\phi}{\hat H}{\phi}$ along the flow yields
\begin{equation}
    \begin{aligned}
        \dot{E}_{\mathrm{ITE}} &= \expect{\partial_\tau\phi}{\hat{H}}{\phi} + \expect{\phi}{\hat{H}}{\partial_\tau\phi} \\
        &= -\expect{\phi}{\{\bar{H},\hat{H}\}}{\phi} =-2\mu_2 \;,
    \end{aligned}
    \label{sm:eq:ite_rate}
\end{equation}
with the central moments $\mu_m=\langle\bar H^m\rangle$ of the main text. Thus, ITE decreases the energy monotonically at a rate proportional to the instantaneous energy variance, $\mu_2=\operatorname{Var}(\hat H)$.

More generally, in this work we consider normalized evolution generated by a Hermitian operator $\hat G = \hat G ^\dagger$,
\begin{equation}
    \partial_\tau\ket{\phi}
    = -\bigl(\hat G-\avg{\hat G}\bigr)\ket{\phi}
    \equiv -\bar G\ket{\phi}\;,
\end{equation}
and for the centered generators of the main text [Eq.~\maineqref{eq:evo}] $\langle\hat G\rangle=0$. The corresponding energy rate becomes
\begin{equation}
    \label{sm:eq:app}
    \begin{aligned}
        \dot{E}[\hat{G}]
        &= -\bra{\phi}\left\{\hat{H}-E,\hat{G}-\avg{\hat{G}}\right\}\ket{\phi} \\
        &= -2\Rez\bra{\phi}\bar{H}\bar{G}\ket{\phi}\;,
    \end{aligned}
\end{equation}
which by the Cauchy--Schwarz inequality satisfies
\begin{equation}
    -\dot E[\hat G]
    \leq 2\left|\langle\bar H\phi|\bar G\phi\rangle\right|
    \leq 2\sqrt{\operatorname{Var}(\hat H) \operatorname{Var}(\hat G)}\;.
\end{equation}
In fact, ITE with $\hat G=\bar H$ saturates this bound with the steepest instantaneous energy descent at fixed generator variance $\mu_2$. AITE~\cite{benavides_riveros_augmentingimaginarytimeevolution_2026}, discussed in Sec.~\ref{sm:aite}, achieves a faster instantaneous descent by relaxing this constraint, additionally taking the third moment $\mu_3$ into account.

% ============================================================================

\subsection{Ground-state convergence}
\label{sm:ite:iconvergence}

Let $\hat H\ket{E_j}=E_j\ket{E_j}$, with a unique ground state, $E_0<E_1\leq\cdots\leq E_{\mathcal{D}-1}\equiv E_{\max}$, and a Hilbert-space dimension $\mathcal{D} \equiv \dim\mathcal H$. We denote the excitation energies $\Delta_j=E_j-E_0$, such that the gap $\Delta=\Delta_1$, the bandwidth $W=\Delta_{\mathcal D-1}$, and the energy error is $\epsilon(\tau)=E(\tau)-E_0$, with initial value $\epsilon_0=\epsilon(0)$.

For an arbitrary normalized state $\ket{\phi}=\sum_j c_j\ket{E_j}$, where the occupation probabilities are $\smash{p_j=|c_j|^2\in[0,1]}$ and $\smash{c_j \equiv \langle E_j | \phi \rangle}$ are the eigenbasis overlaps, the energy error is written as
\begin{equation}
    \epsilon=\sum_{j=0}^{\mathcal D-1}p_j\Delta_j\;.
\end{equation}
Evolving the unnormalized state from the initial $\ket{\psi(0)}=\ket{\phi}$ gives
\begin{equation}
    \ket{\psi(\tau)} = \sum_j c_j e^{-E_j\tau}\ket{E_j}\;,
    \label{sm:eq:stateevo}
\end{equation}
so that each excited-state contribution is exponentially suppressed relative to $\ket{E_0}$,
\begin{equation}
    \frac{\langle E_j|\psi(\tau)\rangle}{\langle E_0|\psi(\tau)\rangle}
    = \frac{c_j}{c_0}\,e^{-\Delta_j\tau}\;,
    \quad j>0\;,
    \label{sm:eq:sm:ite_amplitude_decay}
\end{equation}
provided $c_0\neq0$. The asymptotic rate is set by the smallest excitation energy with nonzero initial overlap,
\begin{equation}
    \Delta_* = \min_{j>0:\, c_j\neq0}\Delta_j \geq\Delta\;,
\end{equation}
dubbed the accessible gap. Symmetry restrictions may remove lowest-lying excitations, effectively resulting in $\Delta_*>\Delta$.

Finally, using $\Delta_j^2\geq\Delta_* \Delta_j$ for $j>0$, one relates the energy variance $\mu_2$ to the energy error $\epsilon$
\begin{equation}
    \mu_2=\sum_{j>0}p_j\Delta_j^2-\epsilon^2 \geq \epsilon(\Delta_* - \epsilon)\;,
    \label{sm:eq:mu2bound}
\end{equation}
which combined with Eq.~\eqref{sm:eq:ite_rate} yields,
\begin{equation}
    \dot\epsilon_{\rm ITE} \leq -2\epsilon(\Delta_* - \epsilon)\;,
    \label{sm:eq:sm:ite_gap_flow}
\end{equation}
which guarantees a decay rate of at least $2(\Delta_*-\epsilon)$ and by Eq.~\eqref{sm:eq:sm:ite_amplitude_decay}, the asymptotic rate is exactly $2\Delta_*$, such that the convergence of ITE slows down for a small accessible gap.

%%%%%%%%%%%%%%%%%%%%%%%%%%%%%%%%%%%%%%%%%%%%%
% TWO LEVEL
%%%%%%%%%%%%%%%%%%%%%%%%%%%%%%%%%%%%%%%%%%%%%

\subsection{Two-level asymptotics}
\label{sm:two_level_intro}

Consider the asymptotic two-level limit, where the dynamics is governed by the ground state $\ket{E_0}$ and the lowest accessible excitation $\ket{E_*}$, with $E_*=E_0+\Delta_*$. Up to relative phase,
\begin{equation}
    \lvert \phi^{(2)} \rangle \simeq \sqrt{1-p}\ket{E_0} + \sqrt{p}\ket{E_*}\;, \quad p=\frac{\epsilon}{\Delta_*}\;,
    \label{sm:eq:proba_two_level}
\end{equation}
and $p \ll 1$. Notably, however, within the two-level subspace the relations below are exact for any $p\in[0,1]$. Since $E = (1- p)E_0 + p E_* = E_0 + p\Delta_*$, the centered Hamiltonian $\bar H$, \ie ITE generator, has eigenvalues $g^{\rm ITE}_{0} = -p\Delta_*$, $g^{\rm ITE}_{*} = (1-p)\Delta_*$. The central moments $\mu_m$,
\begin{equation}
    \mu_m = (1-p)(- p\Delta_*)^m + p[(1-p)\Delta_*]^m\;,
    \label{sm:eq:twolevel_central_moments}
\end{equation}
depend only on the current state. In particular, for $m=2,3$, and then substituting $p=\epsilon/\Delta_*$ [Eq.~\eqref{sm:eq:proba_two_level}], the relevant moments are
\begin{equation}
    \begin{aligned}
    \mu_2 &= \Delta_*^2p(1-p) = \Delta_*\epsilon-\epsilon^2\;, \\
    \mu_3 &= \Delta_*^3p(1-p)(1-2p) = \Delta_*^2\epsilon - 3\Delta_*\epsilon^2+2\epsilon^3\;,
    \end{aligned}
    \label{sm:eq:twolevel_moments}
\end{equation}
yielding the energy skewness~\cite{pearson_contributionsmathematicaltheory_1895} of the main text
\begin{equation}
    \kappa=\frac{\mu_3}{\mu_2^{3/2}} = \frac{1-2p}{\sqrt{p(1-p)}}
    \simeq\sqrt{\frac{\Delta_*}{\epsilon}}
    \quad (\epsilon\to0)\;.
    \label{sm:eq:twolevel_skewness}
\end{equation}
Introducing the excited-to-ground-state amplitude ratio
\begin{equation}
    r(\tau) \equiv \left|\frac{\langle \phi ^{(2)} | E_* \rangle}{\langle \phi ^{(2)} | E_0 \rangle}\right| = \sqrt{\frac{p}{1-p}}\;,
    \label{sm:eq:excited_to_ground_ratio}
\end{equation}
and solving for $p$, one obtains 
\begin{equation}
    p(\tau)=\frac{r^2}{1+r^2}\;, \quad \epsilon(\tau)=\Delta_*p =\frac{\Delta_*r^2}{1+r^2}\simeq \Delta_*r^2\;,
    \label{sm:eq:twolevel_energy_ratio}
\end{equation}
which hold for any evolution.

For ITE, following Eq.~\eqref{sm:eq:sm:ite_amplitude_decay}, $r(\tau)=r(0)\exp(-\Delta_*\tau)$, and differentiating Eq.~\eqref{sm:eq:twolevel_energy_ratio} gives
\begin{equation}
    \dot\epsilon _{\rm ITE} = -\frac{2\Delta_*^2r^2}{(1+r^2)^2}=-2\epsilon(\Delta_*-\epsilon)=-2\mu_2\;,
    \label{sm:eq:twolevel_flow_saturated}
\end{equation}
where we used $\smash{\mu_2=\epsilon(\Delta_*-\epsilon)=\Delta_*^2p(1-p)}$ [Eq.~\eqref{sm:eq:twolevel_moments}] and $p(1-p)=r^2/(1+r^2)^2$. The bound of Eq.~\eqref{sm:eq:sm:ite_gap_flow} is thus saturated, \ie ITE is as slow as the gap permits. For $\epsilon \ll \Delta_*$ the rate reduces to $\dot\epsilon_{\rm ITE}\simeq-2\Delta_*\epsilon$, and therefore $\epsilon_{\rm ITE} \simeq \Delta_*r(0)^2\exp(-2\Delta_*\tau)$.

%%%%%%%%%%%%%%%%%%%%%%%%%%%%%%%%%%%%%%%%%%%
% DOUBLE BRACKET
%%%%%%%%%%%%%%%%%%%%%%%%%%%%%%%%%%%%%%%%%%%

\subsection{Double-bracket formulation}
\label{sm:ite_doublebracket}

For the pure-state density matrix $\hat\rho=\ket{\phi}\bra{\phi}$, Eq.~\eqref{sm:eq:sm:ite_norm} is equivalent to the double-bracket flow~\cite{brockett1991,bloch_completelyintegrablegradient_1992,xiaoyue_strategiesoptimizingdoublebracket_2024, Zander2026,gluza_doublebracketquantumalgorithms_2026},
\begin{equation}
    \partial_\tau\hat\rho = -\{\bar H,\hat\rho\} = -[\hat\rho,[\hat\rho,\hat H]] = -i[\hat\Omega_{\rm ITE},\hat\rho]\;,
    \label{sm:eq:sm:doublebracket}
\end{equation}
with the Hermitian tangent operator $\hat\Omega_{\rm ITE}\equiv-i[\hat H,\hat\rho]$. The same holds for any Hermitian generator $\hat G$ of Eq.~\maineqref{eq:evo}, with $\hat\Omega_G\equiv-i[\hat G,\hat\rho]$ and $\partial_\tau\ket{\phi}=-i\hat\Omega_G\ket{\phi}=-\bar G\ket{\phi}$. The energy rate and the norm of $\hat\Omega_G$ are
\begin{equation}
    \dot E[\hat G]=-\langle\hat\Omega_G,\hat\Omega_{\rm ITE}\rangle_{\rm HS}\;,
    \quad
    \lVert \hat\Omega_G \rVert_{\rm HS}^2 = 2\operatorname{Var}(\hat G)\;,
    \label{sm:eq:omega_g_rate}
\end{equation}
with the Hilbert--Schmidt inner product $\langle A,B\rangle_{\mathrm{HS}}=\Tr(A^\dagger B)$, where the former reproduces Eq.~\eqref{sm:eq:app}. Fixing the Hilbert--Schmidt norm of $\smash{\hat\Omega_G}$ is thus equivalent to fixing the variance of $\hat G$. By the Cauchy--Schwarz inequality, ITE is then the direction of fastest instantaneous energy decrease~\cite{alghadeer_doublebracketalgorithmiccooling_2025a}, as in Sec.~\ref{sm:ite:norm}.

% -------------------------------------------
% AITE
% -------------------------------------------

\section{Augmented imaginary-time evolution}
\label{sm:aite}

The AITE evolution replaces the pure-state density matrix $\hat \rho$ with a rank-one projector $\hat{\mathcal{O}} \in \mathcal{P}_1(\mathcal{H})$, where
\begin{equation}
    \mathcal{P}_1(\mathcal{H}) \equiv \{\hat{\mathcal{O}}=\hat{\mathcal{O}}^\dagger:\hat{\mathcal{O}}^2=\hat{\mathcal{O}},\ \Tr [\hat{\mathcal{O}}] = 1\}\;,
\end{equation}
denotes the set of all pure-state density operators on the Hilbert space $\mathcal{H}$, thereby generalizing the double-bracket formulation~\cite{benavides_riveros_augmentingimaginarytimeevolution_2026}. Accordingly, the ITE tangent operator $\hat\Omega_{\rm ITE}$ of Eq.~\eqref{sm:eq:sm:doublebracket} is replaced by $\hat \Omega _{\mathcal{O}} = -i[\hat H,\hat{\mathcal O}]$, yielding
\begin{equation}
    \partial_\tau\ket{\phi} = -i\hat\Omega_{\mathcal O}\ket{\phi}\;,
    \label{sm:eq:aite_state}
\end{equation}
and the energy rate is linear in the projector $\hat{\mathcal{O}}$,
\begin{equation}
    \dot E = \Tr\!\left[\hat\chi\,\hat{\mathcal O}\right]\;, \quad \hat\chi=-i[\hat H,\hat\Omega_{\rm ITE}]\;.
    \label{sm:eq:rate_aite}
\end{equation}

AITE searches for the augmented projector $\hat\rho_A$ that minimizes Eq.~\eqref{sm:eq:rate_aite} over $\mathcal{P}_1$~\cite{benavides_riveros_augmentingimaginarytimeevolution_2026}, restricted to the second-order Krylov space $\mathcal S_H=\mathrm{span}\{\ket{\phi},\ket{v}\}$, with $\ket{v}=\bar H\ket{\phi}/\sqrt{\mu_2}$. In this subspace,
%the matrix elements of $\hat \chi$ are obtained from $\bra{\phi}\hat\chi\ket{\phi}=-2\mu_2$, $\bra{v}\hat\chi\ket{v}=2\mu_2$, and $\bra{\phi}\hat\chi\ket{v}=-\mu_3/\sqrt{\mu_2}$, which yield
%
\begin{equation}
    \chi \big|_{\mathcal S_H}
    =-2\mu_2
    \begin{pmatrix}
        1 & \kappa /2\\ \kappa / 2 & -1
    \end{pmatrix}\;,
    \label{sm:eq:chi_krylov}
\end{equation}
with the eigenvalues $\smash{\lambda_\pm=\pm2\mu_2\sqrt{1+\kappa^2/4}}$. Using the lowest-eigenvalue eigenvector
\begin{equation}
    \ket{\lambda_-}=\cos\varphi \ket{\phi}+\sin\varphi \ket{v}\;,
    \label{sm:eq:aite_eigenvector}
\end{equation}
where $\tan(2\varphi)=\kappa/2$ and $2\varphi\in(-\pi/2,\pi/2)$, the augmented projector therefore becomes $\hat\rho_A=\ket{\lambda_-}\bra{\lambda_-}$. Since $\Tr[\hat\chi\hat\rho_A]=\lambda_-$, the energy rate in Eq.~\eqref{sm:eq:rate_aite} gives Eq.~\maineqref{eq:aite_rate} from the main text,
\begin{equation}
    \dot E_{\rm AITE}=-2\mu_2\sqrt{1+\kappa^2/4}\;\le\;-2\mu_2=\dot E_{\rm ITE}\;,
    \label{sm:eq:rate_aite_min}
\end{equation}
with strict inequality for $\kappa \neq 0$, \ie for a skewed instantaneous energy distribution~\cite{benavides_riveros_augmentingimaginarytimeevolution_2026}.

Inserting $\hat{\mathcal O}=\hat\rho_A$ into Eq.~\eqref{sm:eq:aite_state} gives $-i\hat\Omega_{\rho_A}\ket{\phi}=-\hat G_{\rm AITE}\ket{\phi}$, \ie the evolution generated by $\hat G_{\rm AITE}$ [Eq.~\eqref{sm:eq:aite_generator}],
\begin{equation}
    \hat G_{\rm AITE}=a\bar H+b\,(\bar H^2-\mu_2)\;,
    \label{sm:eq:aite_generator}
\end{equation}
where $\smash{a=\cos(2\varphi)}$ and $\smash{b=\sin(2\varphi)/(2\sqrt{\mu_2})}$ and the constant $-b\mu_2$ centers the generator $\smash{\langle\hat G_{\rm AITE}\rangle=0}$. 

Although Eq.~\eqref{sm:eq:rate_aite_min} holds for either sign of $\kappa$, we use $\hat G_{\rm AITE}$ only when $\mu_2>0$ and $\mu_3>0$ (\ie $b>0$), as stochastic evolution is less stable otherwise. In all other cases we take the ITE step, $(a,b)=(1,0)$. This fallback is rarely triggered for the models considered.

% -------------------------------------------
% Asymptiotic
% -------------------------------------------

\subsection{Two-level AITE asymptotics} \label{sm:two_level_aite}

Returning to the two-level picture, Eqs.~\eqref{sm:eq:twolevel_moments} and \eqref{sm:eq:twolevel_skewness} give $1+\kappa^2/4=[4p(1-p)]^{-1}$, and the energy rate in Eq.~\eqref{sm:eq:rate_aite_min} becomes
\begin{equation}
    \dot\epsilon_{\rm AITE} = -\Delta_*\sqrt{\epsilon(\Delta_*-\epsilon)}\;\simeq\;-\Delta_*^{3/2}\sqrt{\epsilon}\;,
    \label{sm:eq:aite_asymptotic_rate}
\end{equation}
in contrast to $\dot\epsilon_{\rm ITE}$ that is linear in $\epsilon$, \ie $\epsilon_{\rm ITE}\propto e^{-2\Delta_*\tau}$. Integrating Eq.~\eqref{sm:eq:aite_asymptotic_rate} with $\epsilon=\Delta_*\sin^2\vartheta$ gives $\dot\vartheta=-\Delta_*/2$ and
\begin{equation}
    \epsilon_{\rm AITE} (\tau) = \Delta_*\sin^2\!\left[\arcsin\sqrt{\frac{\epsilon_0}{\Delta_*}}-\frac{\Delta_*\tau}{2}\right]\;,
    \label{sm:eq:aite_asymptotic_flow}
\end{equation}
and remarkably, the error vanishes at the finite time $\tau_*=(2/\Delta_*)\arcsin\sqrt{\epsilon_0/\Delta_*}\simeq 2\sqrt{\epsilon_0/\Delta_*^3}$ ($\epsilon_0\ll\Delta_*$), with $\epsilon_{\rm AITE}\simeq(\Delta_*^3/4)(\tau_*-\tau)^2$ as $\tau\to\tau_*$, whereas ITE converges only asymptotically. This concerns the continuous flow in the two-level limit, where higher excitations are neglected. The discrete updates are analyzed later in Sec.~\ref{sm:two_level_euler}.

With $\mu_2=\Delta_*^2p(1-p)$ [Eq.~\eqref{sm:eq:twolevel_moments}] and $\kappa=(1-2p)/\sqrt{p(1-p)}$ [Eq.~\eqref{sm:eq:twolevel_skewness}], $\tan 2\varphi=\kappa/2$ gives $a=2\sqrt{p(1-p)}$ and $b=(1-2p)/(2\Delta_*\sqrt{p(1-p)})$, so the centered generator $\smash{\hat G_{\rm AITE}}$ has the two-level eigenvalues
\begin{equation}
    g^{\rm AITE}_{0}=-\frac{\Delta_*r}{2}\;, \quad g^{\rm AITE}_{*}=\frac{\Delta_*}{2r}\;.
    \label{sm:eq:twolevel_aite_eigs}
\end{equation}
%

% -------------------------------------------
% Hamiltonian Bandwidth
% -------------------------------------------

\section{Hamiltonian bandwidth bound}\label{sm:hamiltonian_bw}

The 
%\DIFaddbegin \DIFadd{ground-state asymptotic }\DIFaddend 
\expl Euler stability conditions [Eq.~\maineqref{em:eq:stability}] are expressed in terms of the generator bandwidth $W_G=\max_j g_j-g_0$, with $g_j$ the eigenvalue of $\smash{\hat G}$ for the eigenstate $\ket{E_j}$ of $\hat H$. For ITE, $W_G=W$ is the bandwidth of $\hat H$. For AITE, $g_j=a(E_j-E)+b[(E_j-E)^2-\mu_2]$ and 
%\DIFaddbegin \DIFadd{, in this regime, }\DIFaddend 
$W_G=W_A=W[a+b(W-2\epsilon)]$ [Eq.~\maineqref{em:eq:aite_bandwidth}]. 

Both $W$ and $W_A$ are generally unavailable for system sizes beyond exact diagonalization. For spin-$1/2$ systems, a conservative bound follows directly from the Hamiltonian upon expanding $\hat H$ in the Pauli strings $\hat P_\gamma=\bigotimes_{\ell=1}^N\hat\sigma^{\gamma_\ell}_\ell$, with $\gamma_\ell\in\{0,x,y,z\}$, $\smash{\hat\sigma^0=\hat I}$. The Pauli strings are Hermitian, satisfying $\hat P_\gamma^2=\hat I$, $\smash{\Tr[\hat P_\gamma\hat P_{\gamma'}]=2^N\delta_{\gamma\gamma'}}$ and
\begin{equation}
    \hat H = h_0\hat I + \sum_{\gamma\neq0} h_\gamma \hat P_\gamma \;,
    \quad h_\gamma = 2^{-N}\Tr(\hat P_\gamma\hat H)\;.
\end{equation}
The identity contribution $\smash{\hat P_0=\hat I}$ does not affect spectral differences and may be discarded. With $\lVert\cdot\rVert_{\rm op}$ the operator norm, $\smash{\lVert\hat P_\gamma\rVert_{\rm op}=1}$, and the triangle inequality gives
\begin{equation}
    W \leq 2\left\lVert \hat H-h_0\hat I \right\rVert_{\rm op} \leq  2\sum_{\gamma\neq0}|h_\gamma| \equiv \overline W \;,
    \label{app:eq:pauli_bandwidth_bound}
\end{equation}
which might be loose depending on the model. Replacing $W \rightarrow \overline W \geq W$ in Eq.~\maineqref{em:eq:stability}~\cite{hairer_solvingordinarydifferential_1991} gives the computable step $\delta\tau_{\rm ITE}<2/\overline W$.

For AITE with $a,b \geq 0$, Eq.~\maineqref{em:eq:aite_bandwidth} gives $W_A=W[a+b(W-2\epsilon)]\leq a\overline W+b\overline W^2\equiv\overline W_A$. In practice, we use the tighter range $\max f-\min f$ of $f(x)=ax+b(x^2-\mu_2)$ over $\smash{x\in[h_0-\overline W/2-E,\,h_0+\overline W/2-E]}$, which never exceeds $\overline W_A$. Note that $\overline W$ depends only on $\hat H$, while $a$ and $b$ come from the energy moments AITE already computes, so no further spectral information is needed.

% -------------------------------------------
% Variational
% -------------------------------------------

\section{Variational Monte Carlo and neural quantum states} \label{sm:nqs}

Neural quantum states (NQS)~\cite{carleo_solvingquantummanybody_2017} realize a powerful neural-network representation of a many-body wavefunction within the variational Monte Carlo (VMC) method~\cite{foulkes_quantummontecarlo_2001, becca_quantummontecarlo_2017}. We describe the NQS ansatz [Sec.~\ref{sm:ansatz}] and Monte Carlo estimators [Sec.~\ref{sm:stochastic}] used for the stochastic evolution throughout this work. We use NetKet~\cite{netket3:2022,netket2:2019} for Hilbert-space representations and sampling, Flax~\cite{flax2020github} for neural network architectures, and JAX~\cite{jax2018github} for all other numerical operations.
%Importantly, all computations are performed in double precision.

We use the TFI [Eq.~\maineqref{eq:ising}] and $J_1$-$J_2$ [Eq.~\maineqref{eq:j1j2}] models of the main text on the $L_x\times L_y$ square lattice with $N=L_xL_y$ sites and periodic boundary conditions (PBC), \ie $\hat\sigma^{\nu}_{(L_x+1,\ell_y)}=\hat\sigma^{\nu}_{(1,\ell_y)}$ and $\hat\sigma^{\nu}_{(\ell_x,L_y+1)}=\hat\sigma^{\nu}_{(\ell_x,1)}$ for $\nu=x,y,z$, where $\ell=(\ell_x,\ell_y)$. For the symmetry-adapted ans\"atze, we restrict $\Psi_{\btheta}$ to the translation-invariant sector with total quasimomentum $\mathbf k=\mathbf 0$, which has dimension $\mathcal D\sim2^N/N$.

The TFI model has additional $\mathbb{Z}_2$ spin-parity symmetry $\hat\Pi=\prod_\ell\hat\sigma^x_\ell$, which commutes with lattice translations. For the symmetry-adapted RBM, we restrict $\Psi_{\btheta}$ to the even sector $\hat\Pi=+1$, \ie $\Psi_{\btheta}(-\mathbf s)=\Psi_{\btheta}(\mathbf s)$. Together with $\mathbf k=\mathbf 0$, this gives $\mathcal{D}_{\rm TFI} \sim 2^N / (2N)$. The ViT runs of the main text are not parity projected. No other symmetries are imposed.

The $J_1$-$J_2$ model has a global $U(1)$ symmetry, \ie it conserves the total magnetization $\hat S^z_{\rm tot}=\sum_\ell\hat S^z_\ell$. We restrict $\Psi_{\btheta}$ to the $S^z_{\rm tot}=0$ sector for all ans\"atze. Together with $\mathbf k=\mathbf 0$, this gives $\mathcal{D}_{J_1 J_2} \sim \binom{N}{N/2}/N$. For even $L_x$ and $L_y$, the square lattice is bipartite with sublattices $A$ and $B$. We apply the Marshall sign transformation~\cite{marshall_antiferromagnetism_1955}, $\hat U_{\rm M} = \prod_{\ell\in B}\hat\sigma^z_\ell$. At $J_2=0$, the transformed ground-state can be chosen non-negative in the $S^z_{\rm tot}=0$ sector. For $J_2>0$, this sign structure is no longer exact~\cite{richter_violationmarshallpeierlssign_1994}.

\subsection{Variational ansatz}
\label{sm:ansatz}

We use the wavefunction $\Psi_{\btheta}(\mathbf s)$, the $\smash{\hat\sigma^z}$ product basis, and the logarithmic derivatives $O_k$ of the main text.

For exact summation and small systems, we consider the single-layer complex restricted Boltzmann machines (RBM)~\cite{carleo_solvingquantummanybody_2017,nomura_helpingrestrictedboltzmann_2021} ans\"atze with $M=\alpha N$ hidden units and hidden-unit density $\alpha=M/N \in \mathbb Z_{>0}$. Tracing out the hidden variables $h_m=\pm1$ yields
\begin{equation}
    \begin{aligned}
        \Psi_{\btheta}^{\rm RBM}(\mathbf s) &= \sum_{\mathbf h}\exp\Bigl(\sum_\ell a_\ell s_\ell + \sum_m h_m\mathcal F_m(\mathbf s) \Bigr) \\
        &= \exp\Bigl(\sum_\ell a_\ell s_\ell\Bigr) \prod_{m=1}^{M}2\cosh\mathcal F_m(\mathbf s)\;,
    \end{aligned}
    \label{sm:eq:rbm}
\end{equation}
where $\mathcal F_m(\mathbf s)=b_m+\sum_\ell W_{m\ell}s_\ell$ and $\btheta=\{a_\ell,b_m,W_{m\ell}\}$ represents $N_p = N+M+NM = N+\alpha N(N+1)$ complex variational parameters. Increasing $\alpha$ enlarges the ansatz capacity, while the achievable ground-state representation accuracy depends on the Hamiltonian and imposed symmetries~\cite{carleo_solvingquantummanybody_2017, reh_optimizingdesignchoices_2023, reinhart_goingdeepreaching_2026a}. Logarithmic derivatives of the RBM ansatz admit closed-form analytical expressions~\cite{carleo_solvingquantummanybody_2017}.

To account for translation symmetry, we further employ a symmetry-adapted RBM~\cite{choo_symmetriesmanybodyexcited_2018,nomura_helpingrestrictedboltzmann_2021,netket3:2022}, with parameters shared across lattice translations, reducing the number of independent complex parameters to $N_p=\alpha(N+1)+1$. For the TFI model, we additionally enforce even spin parity, $\Psi_{\btheta}(-\mathbf s)=\Psi_{\btheta}(\mathbf s)$, by setting all biases to zero, $a_\ell=b_m=0$, since $\cosh$ is even [Eq.~\eqref{sm:eq:rbm}], which leaves $N_p=\alpha N$. In the $S^z_{\rm tot}=0$ sector of the $J_1$-$J_2$ model, the constraint is enforced directly in the Monte Carlo sampler~\cite{netket3:2022}. Since $\sum_\ell s_\ell=0$ in this sector, the uniform visible bias is redundant, and we omit it, which leaves $N_p=\alpha(N+1)$.

For larger systems or MCMC sampling, we use the vision transformer (ViT) ansatz~\cite{viteritti_transformervariationalwave_2023} with real parameters, where the lattice is split into patches that are embedded and processed by a stack of encoder layers with multi-head, translation-invariant attention. A $\log\cosh$ output returns the complex log-amplitude $\log\Psi_{\btheta}(\mathbf s)$. All ViT runs use $L=4$ encoder layers, embedding dimension $d=16$, and four attention heads, with $2\times2$ patches except on the $4\times4$ lattice, where $1\times1$ patches are used. This gives $N_p=11\,808$ ($4\times4$), $11\,696$ ($6\times4$), $11\,744$ ($6\times6$), and $11\,856$ ($8\times8$) real parameters. Larger networks are left for future studies.

\subsection{Sampling and Monte Carlo estimators}
\label{sm:stochastic}

In VMC, expectation values are evaluated as averages over the Born distribution $p_{\btheta}(\mathbf s)=|\Psi_{\btheta}(\mathbf s)|^2/\langle\Psi_{\btheta}|\Psi_{\btheta}\rangle$. For an operator $\hat A$, the local estimator~\cite{becca_quantummontecarlo_2017} is
\begin{equation}
    A_{\rm loc}(\mathbf s) = \frac{\bra{\mathbf s}\hat A \ket{\Psi_{\btheta}}}{\Psi_{\btheta}(\mathbf s)} = \sum_{\mathbf s'}\langle\mathbf s|\hat A|\mathbf s'\rangle\frac{\Psi_{\btheta}(\mathbf s')}{\Psi_{\btheta}(\mathbf s)}\;,
    \label{sm:eq:local_estimator}
\end{equation}
yielding the expectation value
\begin{equation}
    \langle\hat A\rangle=\sum_{\mathbf s}p_{\btheta}(\mathbf s) A_{\rm loc}(\mathbf s) \equiv\langle A_{\rm loc}\rangle_p\;,
    \label{sm:eq:expectation_sum}
\end{equation}
where $\langle f\rangle_p\equiv\sum_{\mathbf s}p_{\btheta}(\mathbf s)f(\mathbf s)$. Expectation values are approximated by averaging over $N_s$ configurations sampled from $p_{\btheta}$, \ie $\overline f=N_s^{-1}\sum_{i=1}^{N_s}f(\mathbf s_i)$. The centered logarithmic derivatives are $\widetilde O_k=O_k-\overline{O_k}$. 

Samples are generated by Metropolis--Hastings chains with local moves~\cite{becca_quantummontecarlo_2017}. For exact summation, Eq.~\eqref{sm:eq:expectation_sum} is evaluated over the complete basis of the simulated symmetry sector, using the weights $w_i=p_{\btheta}(\mathbf s_i)$~\cite{netket3:2022}.

% -------------------------------------------
% Implicit SR
% -------------------------------------------

\section{Implicit Stochastic Reconfiguration}\label{sm:implicit_least_squares}

We examine the variational interpretation of the \impl Galerkin projection and contrast it with direct minimization of the \impl Euler~\cite{hairer_solvingordinarydifferential_1991} residual $\ket{i^{(n)}}$ norm [Eq.~\maineqref{eq:em:implicit_residual}].

% -------------------------------------------
% Galerkin!
% -------------------------------------------

\subsection{Galerkin projection}

The Galerkin projection $\smash{\langle T_j^{(n)}|i^{(n)}\rangle=0}$, as shown in the main text, yields the \impl SR equation [Eq.~\maineqref{eq:em:sr_implicit}]. In contrast to \expl SR~\cite{sorella_greenfunctionmonte_1998,sorella_wavefunctionoptimization_2005,carleo_solvingquantummanybody_2017}, the tangent-space matrix $K^{(n)}$ accounts for the action on the finite-step tangent shift.

For the Hermitian generator $\hat G_n$, the projected matrix $K^{(n)}$ is also Hermitian, since $\smash{(K_{jk}^{(n)})^*=\langle T_k^{(n)}|\hat G_n|T_j^{(n)}\rangle=K_{kj}^{(n)}}$.
Together with the Hermiticity of $S^{(n)}$, it follows that the projected operator $\mathcal{K}^{(n)} = S^{(n)} + \delta\tau K^{(n)}$ is Hermitian. Consequently, the Galerkin equation $\mathcal{K}^{(n)}\mathbf d_n=-\mathbf F^{(n)}$ is equivalent to the stationarity condition of a real quadratic functional, consistent with the Ritz--Galerkin approach. Specifically
\begin{equation}
    \mathcal Q_n\left(\mathbf d_n\right) = \mathbf d_n^\dagger \mathcal{K}^{(n)}\mathbf d_n +2\Re(\mathbf d_n^\dagger\mathbf F^{(n)})\;,
    \label{sm:eq:galerkin_quadratic}
\end{equation}
gives, by Wirtinger calculus in $\mathbf d_{n}^*$, the stationarity condition
\begin{equation}
    \frac{\partial\mathcal Q_n}{\partial\mathbf d_n^*} = \mathcal{K}^{(n)}\mathbf d_n+\mathbf F^{(n)}=0\;,
\end{equation}
which reproduces Eq.~\maineqref{eq:em:sr_implicit} of the main text. Completing the squared norm and identifying the \expl McLachlan residual gives
\begin{equation}
    \begin{aligned}
        \mathcal Q_n 
        &=\overbrace{\left\| \sum_k d_{n,k}\ket{T_k^{(n)}} + \hat G_n\ket{\Phi_n} \right\|^2}^{\mathcal R(\mathbf d_n)} \\
        & + \delta\tau\,\mathbf d_n^\dagger K^{(n)}\mathbf d_n - \langle\hat G_n^2\rangle\;,
    \end{aligned}
    \label{sm:eq:galerkin_functional}
\end{equation}
such that $\mathcal Q_n$ differs from the \expl McLachlan functional~\cite{mclachlan_variationalsolutiontimedependent_1964} $\mathcal R$ [Eq.~\maineqref{eq:em:residual}] by a finite-step term $\delta\tau\,\mathbf d_n^\dagger K^{(n)}\mathbf d_n$, along with an additive constant independent of $\mathbf d_n$. As $\delta\tau\to0$, the finite-step term vanishes, and the stationarity condition $\smash{\partial_{\mathbf d_n^*}\mathcal Q_n=0}$ recovers \expl SR. The stationary point is a unique minimum provided that $\smash{\mathcal K^{(n)}}$ is positive definite.
%%%%%%%%%%%%%%%%%%
\begin{figure*}[t]
    \centering
    \includegraphics[width=\textwidth]{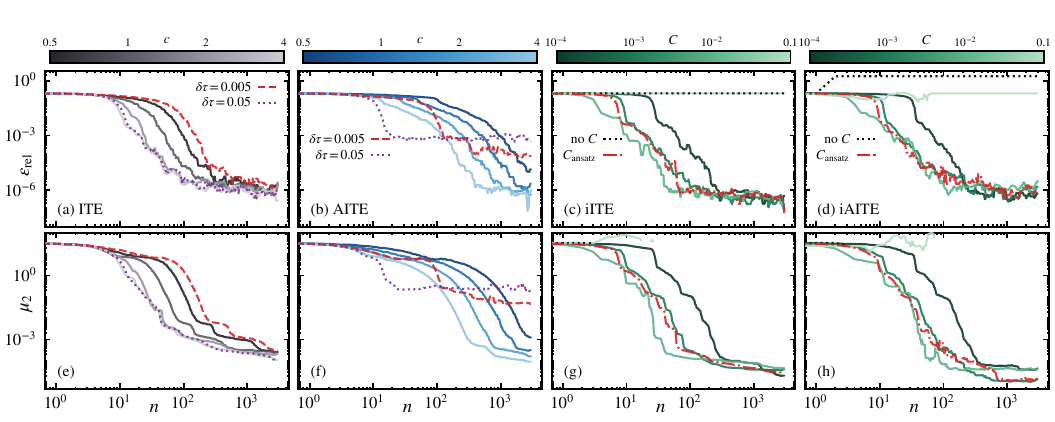}
    \caption{(a)-(d) Relative energy error $\epsilon_{\rm rel}$ and (e)-(h) energy variance $\mu_2$ versus the number of steps $n$ for \expl ITE (a,e), \expl AITE (b,f), \impl ITE (c,g), and \impl AITE (d,h) for the $4\times4$ TFI model at $h=2$ in the even-parity sector, using symmetry-adapted RBMs with $\alpha=8$ and $N_s=4096$ samples per step. Curves are smoothed using an 11-point running median for clarity. (a),(b),(e),(f), colors encode the step-size factor $c\in\{0.5,1,2,4\}$ [Eq.~\eqref{sm:eq:explicit_step}], with $\delta\tau_0=1$, and red dashed and purple dotted curves indicate fixed $\delta\tau=0.005$ and $0.05$, respectively. (c),(d),(g),(h), colors encode the trust-region radius $C\in\{10^{-4},10^{-3},10^{-2},10^{-1}\}$ [Eq.~\eqref{sm:eq:trust_region}], and black dotted and red dash-dotted curves indicate unconstrained evolution (no $C$) and $C_{\rm ansatz}$ [Eq.~\eqref{sm:eq:trust_region}], respectively. The \impl schemes use $\delta\tau_0=1$ and $\tilde c=1$ [Eq.~\eqref{sm:eq:implicit_step_limit}], and in (a),(b),(e),(f) the trust region is fixed at $C=C_{\rm ansatz}$.
    }
    % Curves are smoothed using a sliding window for clarity.}
    \label{sm:fig1}
\end{figure*}
%%%%%%%%%%%%%%%%%%

\subsection{Minimizing the \impl residual}

Alternatively, one may be tempted to minimize the squared \impl Euler residual $\mathcal I(\mathbf d_n)=\|i^{(n)}\|^2$, following McLachlan's least-squares variational principle~\cite{mclachlan_variationalsolutiontimedependent_1964,haegeman_timedependentvariationalprinciple_2011}. For Hermitian $\hat G_n$, its expansion reads
\begin{equation}
    \begin{aligned}
        \mathcal I(\mathbf d_n) &\equiv \lVert i^{(n)}\rVert ^2 = \bra{\Phi_n}\hat G_n^2\ket{\Phi_n} \\
        & + 2\Re\!\left[\sum_k d_{n,k}^*
        \langle T_k^{(n)}|(1+\delta\tau\hat G_n)
        \hat G_n\ket{\Phi_n}\right] \\
        & + \sum_{j,k} d_{n,j}^* d_{n,k}
        \langle T_j^{(n)} |(1+\delta\tau\hat G_n)^2
        |T_k^{(n)}\rangle\;,
    \end{aligned}
    \label{eq:sm:residual_implicit}
\end{equation}
and stationarity with respect to $d_{n,j}^*$ then yields
\begin{equation}
    \begin{aligned}
        0 &= \frac{\partial\mathcal I}{\partial d_{n,j}^*} = \langle T_j^{(n)} |(1+\delta\tau\hat G_n)\hat G_n\ket{\Phi_n} \\
          & + \sum_k d_{n,k}\,\langle T_j^{(n)} |(1+\delta\tau\hat G_n)^2|T_k^{(n)}\rangle\;,
    \end{aligned}
\end{equation}
which, upon expansion becomes,
\begin{equation}
    \begin{aligned}
        0
        &= \underbrace{\langle T_j^{(n)} |\hat G_n
        \ket{\Phi_n}}_{F_j^{(n)}} + \delta\tau \underbrace{\langle T_j^{(n)} |\hat G_n^2 \ket{\Phi_n}}_{\Xi_j^{(n)}} \\
        & + \sum_k d_{n,k}\Biggl[
        \underbrace{\langle T_j^{(n)}|T_k^{(n)}\rangle}
        _{S_{jk}^{(n)}} + 2\delta\tau \underbrace{\langle T_j^{(n)} |\hat G_n |T_k^{(n)}\rangle}_{K_{jk}^{(n)}} \\
        & + \delta\tau^2
        \underbrace{\langle T_j^{(n)} |\hat G_n^2 |T_k^{(n)}\rangle}_{L_{jk}^{(n)}}
        \Biggr] \;,
    \end{aligned}
\end{equation}
and, consequently, with $\mathcal{U}^{(n)} = S^{(n)}+2\delta\tau K^{(n)} + \delta\tau^2 L^{(n)}$
\begin{equation}
    \mathcal{U}^{(n)} \mathbf d_n = - \mathbf F^{(n)}-\delta\tau\mathbf \Xi ^{(n)}\;,
    \label{eq:sm:sr_implicit_damn}
\end{equation}
where the least-squares conditions contain the matrix elements $L^{(n)}$ and $\Xi^{(n)}$ of $\smash{\hat G_n^2}$. These raise the computational cost and the statistical error, in particular for AITE [Eq.~\maineqref{eq:aite_generator}], where $\smash{\hat G_n^2}$ contains $\hat H^4$ terms. The Galerkin projection tests the residual against the tangent vectors $\smash{|T_j^{(n)}\rangle}$, as opposed to $\smash{(1+\delta\tau\hat G_n)|T_j^{(n)}\rangle}$ for the least-squares approach. The two formulations coincide as $\delta\tau \to 0$ and in general differ at finite $\delta\tau$.

% -------------------------------
% \input{SM/sample_space}
% -------------------------------

\section{Euler time-steps}\label{sm:time_steps}

We discretize imaginary-time evolution with adaptive imaginary-time steps $\delta\tau_n=\tau_{n+1}-\tau_n$, such that $\tau_n=\sum_{k=0}^{n-1}\delta\tau_k$. The steps are chosen to control the stability of the \expl and \impl Euler updates with the generator $\smash{\hat G_n}$.

For the \expl schemes, following the main text [Eq.~\maineqref{em:eq:stability}] and Sec.~\ref{sm:hamiltonian_bw}, we use
\begin{equation}
    \delta\tau_n=\min\bigl(\delta\tau_0,\;c\,2/\overline W_G\bigr)\;,
    \label{sm:eq:explicit_step}
\end{equation}
with the guessed step $\delta\tau_0$, step-size factor $c$, and the computable bandwidth bound $\overline W_G\geq W_G$. A factor $c>1$ exceeds the bound $2/\overline W_G$ and is stable only as far as $\overline W_G$ overestimates $W_G$. For ITE, $\overline W_G=\overline W$ [Eq.~\eqref{app:eq:pauli_bandwidth_bound}], while for AITE, $\overline W_G=\overline W_A$ is recomputed at every update from the current $a$, $b$, and $E$.

In the main text, the \expl schemes use $c\in[0.5,1.5]$, and the \impl schemes use $\tilde c=1$ [Eq.~\eqref{sm:eq:implicit_step_limit}] with $\delta\tau_0=1$. Figures~\mainref{fig1} and~\mainref{fig2} use $c=1/2$ for ITE and, in Fig.~\mainref{fig1}, for AITE, \ie $\delta\tau=1/\overline W_G$, with $\delta\tau_0=10^3$, so that the cap $c\,2/\overline W_G$ always applies. 

Further, Fig.~\mainref{fig3} uses $c=1.5$ and $\delta\tau_0=1$ for ITE. In Fig.~\mainref{fig4}, we draw for each scheme the run with the lowest final $E/N$ among the available step controls, which for ITE are $\delta\tau_0=1$ and $c=1.5$. 

For comparison, in Fig.~\ref{sm:fig1}, we instead sweep $c$ over a wider range, and the trust radius $C$ for the \impl schemes, only to check how they affect the runs.

Implicit updates, on the other hand, are A- and L-stable for decaying modes ($g\geq0$)~\cite{hairer_solvingordinarydifferential_1991}. The centered generator, however, includes negative eigenvalues $g_{0,n} < 0$, cf. Sec.~\mainref{em:stiffness}. The corresponding implicit factor $(1+\delta\tau g_{0,n})^{-1}$ becomes singular at $\tilde \delta \tau=-1/g_{0,n}$, changing sign beyond. At each state-dependent $\smash{\hat G_n}$, we restrict the step size to
\begin{equation}
    \delta\tau_n=\min\left(\delta\tau_0, \frac{\tilde c}{\sqrt{2\operatorname{Var}(G_n)}}\right)\;,
    \label{sm:eq:implicit_step_limit}
\end{equation}
with $\delta\tau_0=1$ and $\tilde c=1$ in all figures of the main text and the SM. Furthermore, $g_{0,n}$ can be bound using $\operatorname{Var}(G_n)$. Following Sec.~\ref{sm:ite:iconvergence},
\begin{equation}
    \begin{aligned}
        \operatorname{Var}(G_n) &= \sum_jp_jg_{j,n}^2 = p_0g_{0,n}^2 + \sum_{j>0}p_jg_{j,n}^2 \\
        & \geq p_0g_{0,n}^2+\frac{1}{1-p_0}\Bigl(\sum_{j>0}p_jg_{j,n}\Bigr)^2 \\
        &= p_0g_{0,n}^2+\frac{p_0^2}{1-p_0}\,g_{0,n}^2 = \frac{p_0}{1-p_0}\,g_{0,n}^2\;,
    \end{aligned}
    \label{sm:eq:ground_generator_bound}
\end{equation}
where the second line follows from the Cauchy--Schwarz inequality $(\sum_{j>0}p_jg_{j,n})^2\leq\sum_{j>0}p_j\,\sum_{j>0}p_jg_{j,n}^2$, and the final equality from the centering condition $\sum_jp_jg_{j,n}=0$. The equality holds in the two-level limit and for $p_0>1/3$, Eq.~\eqref{sm:eq:ground_generator_bound} gives $\smash{|g_{0,n}| < [2\operatorname{Var}(G_n)]^{1/2}}$, so the cap of Eq.~\eqref{sm:eq:implicit_step_limit} keeps $1 + \delta\tau_ng_{0,n} > 0$. For AITE, other generator eigenvalues can lie below $g_{0,n}$, so other poles are not excluded. The step controls are compared across all schemes in Fig.~\ref{sm:fig1}.

In addition, to first order, the state changes by $\lVert\ket{\phi_{n+1}}-\ket{\phi_n}\rVert\simeq\delta\tau_n\lVert\hat G_n\ket{\phi_n}\rVert=\delta\tau_n\operatorname{Var}(G_n)^{1/2}$. The cap of Eq.~\eqref{sm:eq:implicit_step_limit} therefore limits the distance moved in the Hilbert space per step to $\tilde c/\sqrt2$. For ITE, $\operatorname{Var}(G_n)=\mu_2 \to 0$ near convergence, recovering $\delta\tau_n=\delta\tau_0$. The cap does not, however, guarantee stability of the projected stochastic evolution, since this argument applies to the deterministic linear problem and does not directly extend to the sampled variational update~\cite{chassagneux_numericalstabilityanalysis_2014}.

\section{Sample-space \expl and \impl estimators} \label{sm:nqs_estimators}

The quantities are evaluated at the current state $\ket{\Phi_n}$, with the generator $\smash{\hat G_n}$ and the step $\delta\tau_n$ [Sec.~\ref{sm:time_steps}], using the estimators of Sec.~\ref{sm:stochastic} and we drop the index $n$ for simplicity.

The QGT and the variational forces [Eq.~\maineqref{eq:em:forces}] are evaluated from the covariance estimators on the same sample set~\cite{becca_quantummontecarlo_2017}
\begin{equation}
    S_{jk} = \langle \widetilde O_j^* \widetilde O_k\rangle_p \;, \quad F_j = \langle\widetilde O_j^* G_{\rm loc}\rangle_p\;,
    \label{sm:eq:sr_estimators}
\end{equation}
and due to centering, $\langle\widetilde O_j\rangle_p = 0$, $F_j$ is invariant under a shift $\hat G \to \hat G+c$.

In addition, the \impl SR formulation requires estimates of the tangent-space matrix $K_{jk}=\langle T_j|\hat G|T_k\rangle$ [Eq.~\maineqref{eq:implicit_sr}]. With $|T_k\rangle = \widetilde O_k|\Phi_n \rangle$ [Eq.~\maineqref{eq:em:tangent_final}], one obtains~\cite{webber_rayleighgaussnewtonoptimizationenhanced_2022, zhang_projectedinverseiteration_2026a, medvidovic_adiabatictransportneural_2026},
% the components of the tangent vectors are $\langle\mathbf s|T_j\rangle=\widetilde O_j(\mathbf s)\Phi_n(\mathbf s)$
% Inserting $\sum_{\mathbf s}\ket{\mathbf s}\bra{\mathbf s}$
%
\begin{equation}
    \begin{aligned}
        K_{jk} &= \sum_{\mathbf s}\widetilde O_j^*(\mathbf s)\,\Phi_n^*(\mathbf s) \langle\mathbf s|\hat G|T_k\rangle \\
        &= \sum_{\mathbf s}|\Phi_n(\mathbf s)|^2 \widetilde O_j^*(\mathbf s) \; \frac{\langle\mathbf s|\hat G|T_k\rangle}{\Phi_n(\mathbf s)} \\
        &= \langle\widetilde O_j^*B_k\rangle_p\;,
    \end{aligned}
    \label{sm:eq:k_estimator}
\end{equation}
where $B_k(\mathbf s) \equiv \langle\mathbf s|\hat G|T_k\rangle/\Phi_n(\mathbf s)$. Differentiating the local estimator in Eq.~\eqref{sm:eq:local_estimator} yields
\begin{equation}
    \frac{\langle\mathbf s|\hat A|T_k\rangle}{\Phi_n(\mathbf s)} = \partial_{\theta_k}A_{\rm loc}(\mathbf s) + \widetilde O_k(\mathbf s) A_{\rm loc}(\mathbf s)\;,
    \label{sm:eq:local_tangent_identity}
\end{equation}
allowing $B_k$ to be evaluated without constructing $|T_k\rangle$. For $\hat A=\hat G_n$, its moments and coefficients are held fixed when differentiating. For  diagonal $\hat A$, the first term vanishes and in general, Eq.~\eqref{sm:eq:local_tangent_identity} involves derivatives of the amplitude ratios $\Psi_{\btheta}(\mathbf s')/\Psi_{\btheta}(\mathbf s)$ over configurations connected by $\hat A$. This might become computationally demanding, especially for AITE [Eq.~\eqref{sm:eq:aite_local_generator}]. 

The local estimators of $\bar H=\hat H-E$ and $\bar H^2$ are
\begin{equation}
    \begin{aligned}
        x(\mathbf s) &= H_{\rm loc}(\mathbf s)-E\;,\\
        y(\mathbf s) &= (H^2)_{\rm loc}(\mathbf s)-2E H_{\rm loc}(\mathbf s)+E^2\;,
    \end{aligned}
    \label{sm:eq:centered_energy_estimators}
\end{equation}
where $(H^2)_{\rm loc}$ uses the product of the Pauli-string expansions of $\hat H$ (Sec.~\ref{sm:hamiltonian_bw}). Since $\hat H$ is Hermitian, the centered moments become
\begin{equation}
    \mu_2=\langle|x|^2\rangle_p\;,\quad
    \mu_3=\Re\langle x^*y\rangle_p\;,\quad
    \mu_4=\langle|y|^2\rangle_p\;,
    \label{sm:eq:moment_estimators}
\end{equation}
which requires powers of $\hat H$ only up to $\hat H^2$. Since $(H^2)_{\rm loc}\neq H_{\rm loc}^2$, $\mu_3$ is not the third central moment of the sampled local energies. 
% Under full summation, the moments are evaluated as vector norms, $\mu_2=\lVert\bar H\Phi_n\rVert^2$, $\mu_3=\Re\langle\bar H\Phi_n|\bar H^2\Phi_n\rangle$, and $\mu_4=\lVert\bar H^2\Phi_n\rVert^2$.

For AITE generator, since $\kappa=\mu_3/\mu_2^{3/2}$ diverges as $\mu_2 \to 0$, we insert $\tan(2\varphi)=\kappa/2$ into the coefficients of Eq.~\eqref{sm:eq:aite_generator} and evaluate them directly from the moments,
\begin{equation}
    a=\frac{2\mu_2^{3/2}}{\varrho}\;, \quad b=\frac{\mu_3}{2\sqrt{\mu_2}\,\varrho}\;,
    \label{sm:eq:stable_aite_coefficients}
\end{equation}
with $\varrho=\sqrt{4\mu_2^3+\mu_3^2}$. For $\mu_3<0$, $b<0$ can yield negative high-energy eigenvalues of  the frozen AITE generator. If $\mu_2\leq0$ or $\mu_3\leq0$ , we fall back to ITE, $(a,b)=(1,0)$.
% If $\mu_3\leq0$ or $\mu_2=0$, we use the ITE generator, $(a,b)=(1,0)$. For $\mu_3\leq0$, $b\leq0$, so the frozen quadratic generator is unbounded from below at high energies, and an \impl step can meet a pole of $(1+\delta\tau\hat G)^{-1}$. Near the ground state $\mu_3>0$ [Eq.~\eqref{sm:eq:twolevel_moments}], so this rule acts only far from it or on noisy estimates.

Finally, using Eq.~\eqref{sm:eq:centered_energy_estimators}, the local estimator of $\hat G_{\rm AITE}$ [Eq.~\maineqref{eq:aite_generator}] and the AITE force yield
\begin{equation}
    \begin{aligned}
        G_{\rm loc}^{\rm AITE} &= ax+b(y-\mu_2)\;,\\
        F_j^{\rm AITE} &= a\langle\widetilde O_j^* x\rangle_p+b\langle\widetilde O_j^* y\rangle_p\;,
    \end{aligned}
    \label{sm:eq:aite_local_generator}
\end{equation}
and variance of the generator can be obtained
\begin{equation}
    \begin{aligned}
        \operatorname{Var}(G_{\rm AITE}) &= \langle\hat G_{\rm AITE}^2 \rangle \\
        &= a^2\mu_2+2ab\mu_3 +b^2(\mu_4-\mu_2^2)\;,
    \end{aligned}
    \label{sm:eq:aite_generator_variance}
\end{equation}
which reduces to $\mu_2$ for ITE. In the numerical implementation, we instead use the sample variance of $G_{\rm loc}^{\rm AITE}$.
%%%%%%%%%%%%%%%%%%
\begin{figure}[!tb]
    \centering
    \includegraphics[width=\columnwidth]{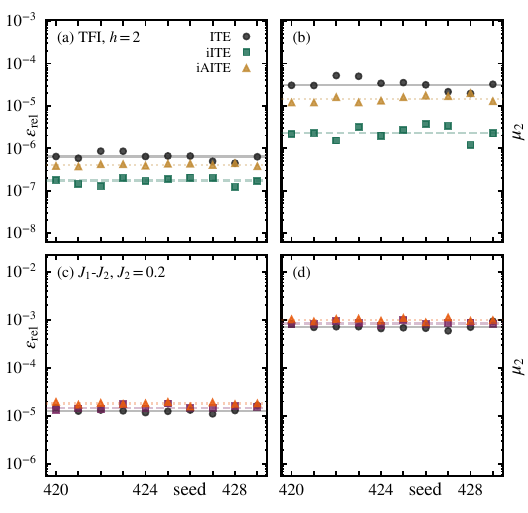}
    \caption{(a),(c) Relative energy error $\epsilon_{\rm rel}$ and (b),(d) energy variance $\mu_2$ of the final $10\%$ of each run versus the random seed, for (a,b) the TFI at $h=2$ and (c,d) the $J_1$-$J_2$ model at $J_2=0.2$, with $\alpha=20$, $N_s=16\,384$. Markers show the median over the final $10\%$ steps.
    }
    \label{sm:fig:seeds}
\end{figure}
%%%%%%%%%%%%%%%%%%

% -------------------------------------------
% REGULARIZATION...
% -------------------------------------------

For $N_s$ samples, the estimators of Eq.~\eqref{sm:eq:sr_estimators} are products of sample matrices. With the $N_s\times N_p$ matrix $X_{ik}=\widetilde O_k(\mathbf s_i)/\sqrt{N_s}$ and the vector $u_i=[G_{\rm loc}(\mathbf s_i)-\overline{G_{\rm loc}}]/\sqrt{N_s}$, where $\overline{G_{\rm loc}}$ is the sample mean, $S=X^\dagger X$ and $\mathbf F=X^\dagger\mathbf u$.

Moreover, in the \impl update, Eq.~\eqref{sm:eq:local_tangent_identity} applied to $\bar H$ and $\bar H^2$, with local estimators $x$ and $y$ [Eq.~\eqref{sm:eq:centered_energy_estimators}], gives the $N_s\times N_p$ matrices
\begin{equation}
    \begin{aligned}
        Y_{ik} &= \frac{\langle\mathbf s_i|\bar H|T_k\rangle}{\sqrt{N_s}\,\Phi_n(\mathbf s_i)} = \frac{[\partial_{\theta_k}x+\widetilde O_k\,x](\mathbf s_i)}{\sqrt{N_s}}\;,\\
        Z_{ik} &= \frac{\langle\mathbf s_i|\bar H^2|T_k\rangle}{\sqrt{N_s}\,\Phi_n(\mathbf s_i)} = \frac{[\partial_{\theta_k}y+\widetilde O_k\,y](\mathbf s_i)}{\sqrt{N_s}}\;.
    \end{aligned}
    \label{sm:eq:implicit_rows}
\end{equation}
Inserting $\hat G=a\bar H+b(\bar H^2-\mu_2)$ into Eq.~\eqref{sm:eq:k_estimator} yields $K=X^\dagger(aY+bZ)-b\mu_2S$. The matrix of the \impl update [Eq.~\maineqref{eq:implicit_sr}]  factorizes as
\begin{equation}
    \begin{aligned}
        S+\delta\tau K &= X^\dagger V\;,\\
        V &= (1-\delta\tau\,b\mu_2)X+\delta\tau(aY+bZ)\;.
    \end{aligned}
    \label{sm:eq:implicit_trial_rows}
\end{equation}

The exact $K$ is Hermitian [Sec.~\ref{sm:implicit_least_squares}]. The same is true for its estimate only under full summation or for $N_s\to\infty$. This is due to $\hat G$ acting only on the ket side of Eq.~\eqref{sm:eq:k_estimator}. Hence, at finite $N_s$, the matrix $X^\dagger V$ is not Hermitian.

\subsection{Solving the linear equations} \label{sm:numerical_details}

The \expl [Eq.~\maineqref{eq:sr}] and \impl [Eq.~\maineqref{eq:implicit_sr}] updates both solve
\begin{equation}
    \bigl(\tilde\lambda I+\mathcal K\bigr)\mathbf d=-\mathbf F\;,
    \label{sm:eq:linear_system}
\end{equation}
where $\mathcal K=S+\delta\tau K=X^\dagger V$ for the \impl update [Eq.~\eqref{sm:eq:implicit_trial_rows}], and $\mathcal K$ is replaced by $S=X^\dagger X$ for the \expl update. The diagonal shift $\tilde\lambda\geq\lambda$, with $\lambda=10^{-3}$, regularizes the system, since the sampled $S$ is singular for $N_p\geq N_s$. We use $\tilde\lambda=\lambda$ unless the trust region below is applied. 

For the \expl update, the matrix of Eq.~\eqref{sm:eq:linear_system} is Hermitian and positive definite, for which Cholesky factorization is used. For the \impl update, the linear system is not Hermitian, and we use an LU factorization~\cite{higham_accuracystabilitynumerical_2002}. For $N_p>N_s$, the system can instead be solved through an $N_s\times N_s$ matrix in sample space~\cite{chen_empoweringdeepneural_2024,rende_simplelinearalgebra_2024}, which reduces the cost of a dense solve from $\mathcal{O}(N_p^3)$ to $\mathcal{O}(N_pN_s^2+N_s^3)$, using
\begin{equation}
    \bigl(\tilde\lambda I+X^\dagger V\bigr)^{-1}=\tilde\lambda^{-1}\bigl[I-X^\dagger\bigl(\tilde\lambda I+VX^\dagger\bigr)^{-1}V\bigr]\;,
    \label{sm:eq:sample_space_inverse}
\end{equation}
with $V$ replaced by $X$ for the \expl update. In this work, we do not use this reduction. For large systems, we use the Arnoldi process, which requires only products with $X$, $V$, and $X^\dagger$, and whose Krylov space does not depend on $\tilde\lambda$~\cite{frommer_restartedgmresshifted_1998}.

%%%%%%%%%%%%%%%%%%
\begin{figure}[!tb]
    \centering
    \includegraphics[width=\columnwidth]{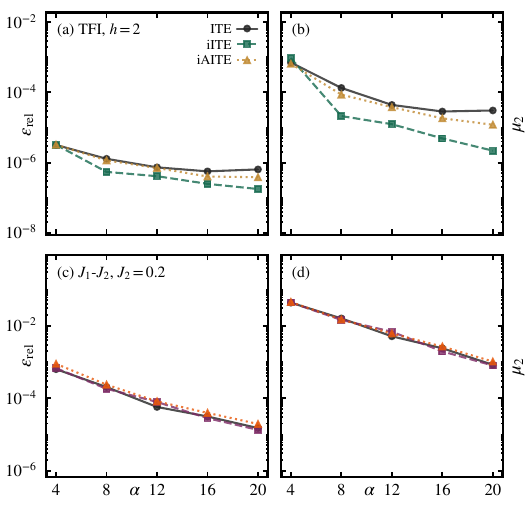}
    \caption{(a),(c) Relative energy error $\epsilon _{\rm rel}$ and (b),(d) energy variance $\mu_2$ of the final $10\%$ of each run versus the RBM width $\alpha$, for (a,b) the TFI at $h=2$ and (c,d) the $J_1$-$J_2$ model at $J_2=0.2$, with $N_s=16\,384$.
    }
    \label{sm:fig:capacity}
\end{figure}
%%%%%%%%%%%%%%%%%%
In addition, we bound the parameter change of each update with a trust region,
\begin{equation}
    \delta\tau^2\lVert\mathbf d\rVert^2\leq C\;,\quad
    C_{\rm ansatz}=10^{-4}\,\frac{\min(n_\theta,2N_s)}{N}\;,
    \label{sm:eq:trust_region}
\end{equation}
with $n_\theta$ real parameters, $N_s$ samples (the number of basis states for full summation), and $N$ lattice sites. The bound applies to the update of every scheme, \expl and \impl. We use $C=C_{\rm ansatz}$ in all runs unless stated otherwise. If the solution at $\tilde\lambda=\lambda$ violates this bound, we increase $\tilde\lambda$ until $\delta\tau\lVert\mathbf d\rVert=\sqrt C$, which is a Levenberg--Marquardt step~\cite{marquardt_algorithmleastsquaresestimation_1963}.
%%%%%%%%%%%%%%%%%%
\begin{figure*}[!t]
    \centering
    \includegraphics[width=\textwidth]{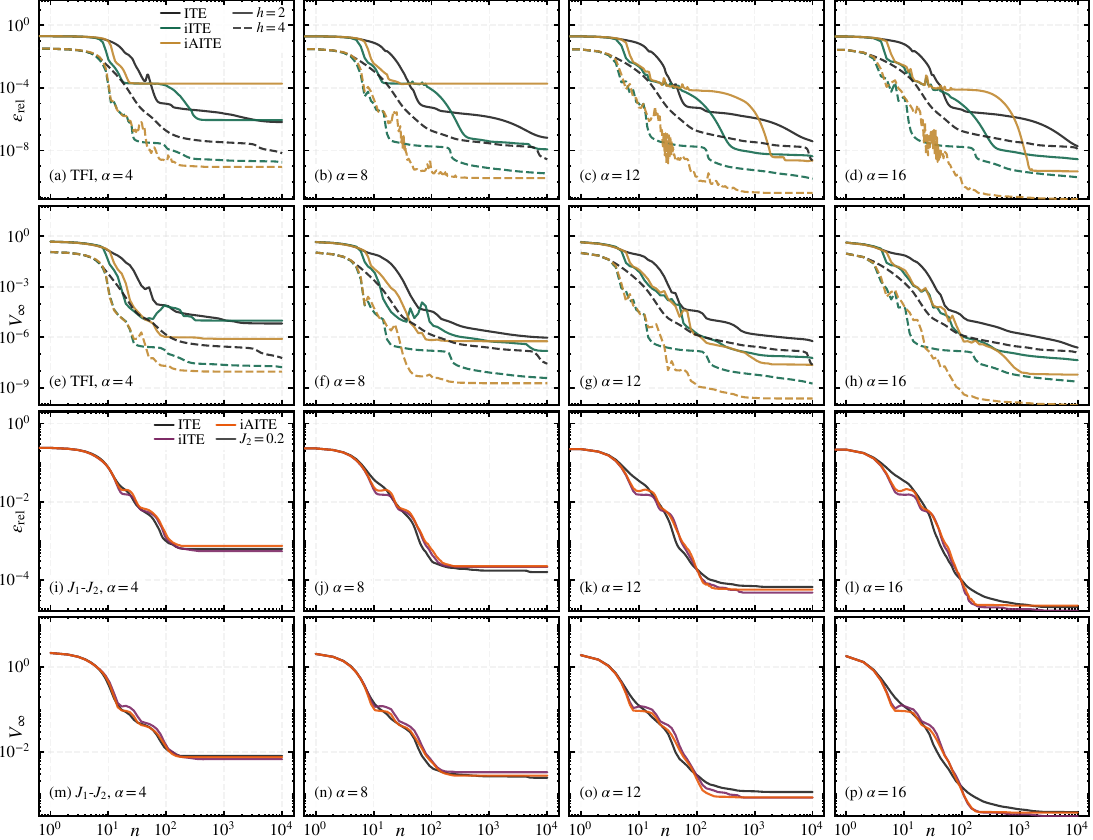}
    \caption{(a)-(d) Relative energy error $\epsilon _{\rm rel}$ and (e)-(h) V-score $V_\infty$ versus the iteration $n$ for the TFI at $h=2$ (solid) and $h=4$ (dashed), and (i)-(l) $\epsilon _{\rm rel}$ (m)--(p) $V_\infty$ for the $J_1$-$J_2$ model at $J_2=0.2$, with full summation over all $2^{16}$ configurations on the $4\times4$ lattice. The columns show the widths $\alpha=4,8,12,16$.
    }
    \label{sm:fig:full}
\end{figure*}
%%%%%%%%%%%%%%%%%%

% -------------------------------------------
% Two-level convergence
% -------------------------------------------

\section{Two-level Euler asymptotics}
\label{sm:two_level_euler}

%%%%%%%%%%%%%%%%%%
\begin{figure}[!tb]
    \centering
    \includegraphics[width=\columnwidth]{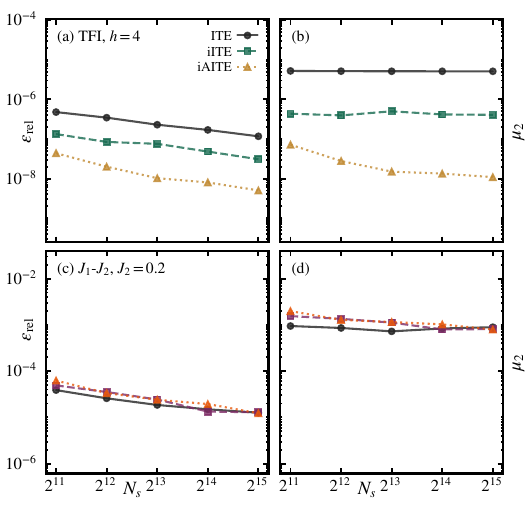}
    \caption{(a),(c) Relative energy error $\epsilon _{\rm rel}$ and (b),(d) energy variance $\mu_2$ of the final $10\%$ of each run versus the number of samples $N_s$ per update, for (a,b) the TFI at $h=4$ and (c,d) the $J_1$-$J_2$ model at $J_2=0.2$, with $\alpha=20$.
    }
    \label{sm:fig:samples}
\end{figure}
%%%%%%%%%%%%%%%%%%
Consider the two-level limit of Secs.~\ref{sm:two_level_intro} and \ref{sm:two_level_aite} at a fixed step $\delta\tau$. For a generator $\smash{\hat G_n}$ frozen at the current state $\ket{\phi_n}$, with eigenvalues $g_{0,n}$ and $g_{*,n}$, the \expl, $r^{(n+1)}$, and \impl, $r'^{(n+1)}$, Euler updates~\cite{hairer_solvingordinarydifferential_1991} of the amplitude ratio give
\begin{equation}
    \begin{aligned}
        r^{(n+1)} 
        & 
        = r^{(n)}\left|\frac{1-\delta\tau g_{*,n}}{1-\delta\tau g_{0,n}}\right|\;,
        \\
        r'^{(n+1)}
        &
        =r'^{(n)}\left|\frac{1+\delta\tau g_{0,n}}{1+\delta\tau g_{*,n}} \right|\;,
    \end{aligned}
    \label{sm:eq:twolevel_euler}
\end{equation}
which hold wherever denominators are nonzero.\par\smallskip

\noindent\textbf{\textit{Imaginary-time evolution.}} For ITE, inserting the eigenvalues $g^{\rm ITE}_{0,n}=-\Delta_*p^{(n)}$ and $g^{\rm ITE}_{*,n}=\Delta_*(1-p^{(n)})$ of Sec.~\ref{sm:two_level_intro}, evaluated at $n$'th step with $p^{(n)}=(r^{(n)})^2/[1+(r^{(n)})^2]$ [Eq.~\eqref{sm:eq:twolevel_energy_ratio}], into Eq.~\eqref{sm:eq:twolevel_euler} and using $\mathcal{T} =\delta\tau\Delta_*$ yields
\begin{equation}
    \begin{aligned}
    r^{(n+1)} & =r^{(n)}\left|
      \frac{1-\mathcal{T} + (r^{(n)})^2}{1+(1+\mathcal{T})(r^{(n)})^2} \right|\;,\\
    r'^{(n+1)} & =r'^{(n)}\left|
      \frac{1+(1-\mathcal{T})(r'^{(n)})^2}{1+\mathcal{T}+(r'^{(n)})^2} \right|\;.
    \end{aligned}
    \label{sm:eq:twolevel_ite}
\end{equation}
Inserting $\epsilon^{(n)} \simeq \Delta _* (r^{(n)})^2$ for $r^{(n)} \rightarrow 0$ at fixed $\delta\tau>0$, the asymptotic recurrences of the \expl $\epsilon^{(n+1)}$ and \impl $\epsilon'^{(n+1)}$ updates are
\begin{equation}
    \begin{aligned}
        \epsilon_{\rm ITE}^{(n+1)} & \sim(1-\mathcal{T})^2\epsilon_{\rm ITE}^{(n)}\;,\\
        \epsilon_{\rm ITE}'^{(n+1)} &\sim\frac{\epsilon_{\rm ITE}'^{(n)}}{(1+\mathcal{T})^2}\;.
    \end{aligned}
    \label{sm:eq:twolevel_ite_error}
\end{equation}
In general, both schemes exhibit linear convergence in $\epsilon^{(n)}$. \Expl ITE requires $0 < \mathcal{T} < 2$, while \impl ITE is, in principle, locally stable for any fixed $\mathcal{T} > 0$. For $\mathcal{T}>2$, however, the excited state is also a stable fixed point of the \impl update, since $r'^{(n+1)}/r'^{(n)}\to|1-\mathcal{T}|>1$ for $r'^{(n)}\to\infty$ [Eq.~\eqref{sm:eq:twolevel_ite}]. Indeed, the \impl ITE step with a frozen generator is an inverse iteration with shift $E_n-1/\delta\tau$, which converges to the eigenstate nearest to the shift~\cite{zhang_projectedinverseiteration_2026a}. A large \impl step can therefore converge to an excited state, depending on the initial state.\par\smallskip

\noindent\textbf{\textit{Augmented imaginary-time evolution.}} For AITE, inserting the eigenvalues of Eq.~\eqref{sm:eq:twolevel_aite_eigs} into Eq.~\eqref{sm:eq:twolevel_euler}, with $\tilde{\mathcal{T}}\equiv \mathcal{T}/2$, gives \expl
\begin{equation}
    r^{(n+1)} =
    r^{(n)}\left|\frac{1-\tilde{\mathcal{T}}/r^{(n)}}{1+ \tilde{\mathcal{T}} r^{(n)}}\right|=\frac{|r^{(n)}-\tilde{\mathcal{T}}|}{1 + \tilde{\mathcal{T}} r^{(n)}}\;,
    \label{sm:eq:twolevel_aite}
\end{equation}
and
\begin{equation}
    \begin{aligned}
        r'^{(n+1)} &= r'^{(n)}\left|\frac{1-\tilde{\mathcal{T}} r'^{(n)}}{1+\tilde{\mathcal{T}}/r'^{(n)}} \right| \\
        &= \frac{(r'^{(n)})^2|1-\tilde{\mathcal{T}} r'^{(n)}|}{r'^{(n)}+\tilde{\mathcal{T}}}\;.
    \end{aligned}
    \label{sm:eq:twolevel_aite_impl}
\end{equation}
As $\smash{r^{(n)} \to 0}$, $\smash{g_{*,n}}$ diverges, so at fixed $\tilde{\mathcal{T}} > 0$ the \expl factor $1-\tilde{\mathcal{T}}/r^{(n)}$ overshoots the excited-state amplitude, giving
\begin{equation}
    \begin{aligned}
        \lim_{r^{(n)} \to 0^+}r^{(n+1)} &= \tilde{\mathcal{T}}\;,\\
        \lim_{r^{(n)} \to 0^+}\epsilon_{\rm AITE}^{(n+1)} &= \frac{\Delta_*\tilde{\mathcal{T}}^2}{1+\tilde{\mathcal{T}}^2}>0\;,
    \end{aligned}
    \label{sm:eq:twolevel_aite_overshoot}
\end{equation}
and thus fixed-step \expl AITE does not converge locally to the ground state. Instead, choosing a decreasing step
$\tilde{\mathcal{T}}_n=cr^{(n)}$, with $0 < c < 2$, yields
\begin{equation}
    \begin{aligned}
        r^{(n+1)} = \frac{|1-c|r^{(n)}}{1+c(r^{(n)})^2}\;,\\
        \epsilon_{\rm AITE}^{(n+1)} \sim(1-c)^2\epsilon_{\rm AITE}^{(n)} \;.
    \end{aligned}
    \label{sm:eq:twolevel_aite_adaptive}
\end{equation}
As a result, a generic stable adaptive step with $\smash{\delta\tau_n \sim (\epsilon_{\rm AITE}^{(n)})^{1/2}}$ again yields linear convergence. In this setting, a tuned $c=1$ removes the excited component in a single iteration.

In contrast, for the \impl update at fixed $\tilde{\mathcal{T}}>0$, the recurrence becomes $r'^{(n+1)} \sim (r'^{(n)})^2/\tilde{\mathcal{T}}$, and hence
\begin{equation}
    \epsilon_{\rm AITE}^{(n+1)} \sim\frac{4}{\delta\tau^2\Delta_*^3} \big(\epsilon_{\rm AITE}^{(n)}\big)^2\;.
    \label{sm:eq:twolevel_quadratic}
\end{equation}
In the two-level limit, \impl AITE therefore exhibits a quadratic, rather than linear, local energy-error convergence without requiring a vanishing time step $\delta \tau$.

The quadratic convergence of Eq.~\eqref{sm:eq:twolevel_quadratic} also holds asymptotically for the full spectrum. In the ground-state regime with $a,b>0$, all higher levels have larger generator eigenvalues than $g_{*,n}$ and are damped faster, so the lowest accessible excitation governs the asymptotics.

% -------------------------------------------
% ADDITIONAL
% -------------------------------------------

\section{Complementing numerical results}
\label{sm:results}

This section collects the results in support of the main text claims. Unless stated otherwise, each run here uses the projected symmetry-adapted RBM on a $4\times4$ lattice with PBC and a diagonal shift $\lambda = 10^{-3}$ using Born sampling. 

We select the TFI at $h=2$ in the even global-$X$-parity sector and the $J_1$-$J_2$ model at $J_2=0.2$ in the $S^z_{\rm tot}=0$ sector. All runs use $\delta\tau_0=1$, with $c=1.5$ for ITE and $\tilde c=1$ for iITE and iAITE (Sec.~\ref{sm:time_steps}), and the trust region $C=C_{\rm ansatz}$ [Eq.~\eqref{sm:eq:trust_region}], unless stated otherwise. In Figs.~\ref{sm:fig:seeds}--\ref{sm:fig:samples}, the relative energy errors $\epsilon _{\rm rel}$ and the energy variances $\mu_2$ are the median over the last $10\%$ of the updates.

\subsection{Seed dependence} 

The NQS optimization depends on the pseudo-random generator seed through the Markov chains and the initial parameters. To quantify this dependence, Fig.~\ref{sm:fig:seeds} repeats one configuration for the seeds $420$--$429$ at $\alpha=20$ and $N_s=16\,384$. The panels Fig.~\ref{sm:fig:seeds}(a),(c) show $\epsilon_{\rm rel}$ and Fig.~\ref{sm:fig:seeds}(b),(d) $\mu_2$ of ITE, iITE and iAITE for each seed, for the TFI Fig.~\ref{sm:fig:seeds}(a),(b) and the $J_1$-$J_2$ model Fig.~\ref{sm:fig:seeds}(c),(d). All seeds show consistent results for the models considered in this work.

\subsection{Ansatz capacity} 

To separate the limits of the ansatz from those of optimization, in Fig.~\ref{sm:fig:capacity} we vary the width of the symmetry-adapted RBM over $\alpha=4,8,12,16,20$ at $N_s=16\,384$ at the same seed. Panels Fig.~\ref{sm:fig:capacity}(a),(c) show $\epsilon_{\rm rel}$ and Fig.~\ref{sm:fig:capacity}(b),(d) $\mu_2$ of ITE, iITE, and iAITE, for the TFI (a),(b) and the $J_1$-$J_2$ model (c),(d). Throughout, enlarging $\alpha$ decreases the final $\epsilon_{\rm rel}$ when fixing the remaining hyperparameters. This is consistent between all models considered here.

In Fig.~\ref{sm:fig:full}, to isolate the effect of the projection onto the RBM manifold, we further sum over the complete basis of the simulated sector of the $4\times4$ lattice, for $\alpha=4,8,12,16$, for the TFI at $h=2$ and $h=4$ and the $J_1$-$J_2$ model at $J_2=0.2$. Panels Fig.~\ref{sm:fig:full}(a)--(d) and Fig.~\ref{sm:fig:full}(i)--(l) show $\epsilon_{\rm rel}$, while panels Fig.~\ref{sm:fig:full}(e)--(h) and Fig.~\ref{sm:fig:full}(m)--(p) show the V-score $V_\infty$, of ITE, iITE, and iAITE as a function of the finite evolution steps $n$, for the TFI [Fig.~\ref{sm:fig:full}(a)--(h)] and the $J_1$-$J_2$ model [Fig.~\ref{sm:fig:full}(i)--(p)]. Each column corresponds to one width $\alpha$.

\subsection{Number of samples} 

Finally, to test how the attainable accuracy depends on the statistical error of the estimators, Fig.~\ref{sm:fig:samples} varies the number of samples over $N_s=2^{11},\ldots,2^{15}$ at $\alpha=20$. Panels Fig.~\ref{sm:fig:samples}(a),(c) show $\epsilon_{\rm rel}$ and Fig.~\ref{sm:fig:samples}(b),(d) $\mu_2$ of ITE, iITE, and iAITE, for the TFI at $h=4$ [Fig.~\ref{sm:fig:samples}(a),(b)] and the $J_1$-$J_2$ model at $J_2=0.2$ [Fig.~\ref{sm:fig:samples}(c),(d)]. While consistent results are shown for both models, the $J_1$-$J_2$ model benefits less from the increased number of samples $N_s$, and the schemes differ less in accuracy than for the TFI model.

\ifdefined\combinedarxiv
    % Bibliography already included in main.tex
\else
    \bibliography{references}
\fi

\fi

\end{document}